\RequirePackage[2020-02-02]{latexrelease}
\documentclass[prb,twocolumn,showpacs,preprintnumbers,superscriptaddress,amsmath,amssymb]{revtex4}
\usepackage[T1]{fontenc}
\usepackage{graphicx}
\usepackage[latin9]{inputenc}
\usepackage{amsmath}
\usepackage{babel}
\usepackage{array}
\usepackage{color}
\usepackage[dvipsnames]{xcolor}

\begin{document}

\title{Implementation of the SCAN Exchange-Correlation Functional with Numerical Atomic Orbitals}

\author{Renxi Liu}
\affiliation{HEDPS, CAPT, College of Engineering, Peking University, Beijing, 100871, P. R. China}
\affiliation{Academy for Advanced Interdisciplinary Studies, Peking University, Beijing, 90871, P. R. China}
\affiliation{AI for Science Institute, Beijing 100080, P. R. China}
\author{Daye Zheng}
\affiliation{AI for Science Institute, Beijing 100080, P. R. China}
\author{Xinyuan Liang}
\affiliation{HEDPS, CAPT, College of Engineering, Peking University, Beijing, 100871, P. R. China}
\affiliation{Academy for Advanced Interdisciplinary Studies, Peking University, Beijing, 90871, P. R. China}
\author{Xinguo Ren}
\affiliation{Beijing National Laboratory for Condensed Matter Physics, Institute of Physics, Chinese Academy of Sciences, Beijing 100190, P. R. China}
\affiliation{Songshan Lake Materials Laboratory, Dongguan, Guangdong 523808, P. R. China}
\author{Mohan Chen}
\affiliation{HEDPS, CAPT, College of Engineering, Peking University, Beijing, 100871, P. R. China}
\affiliation{Academy for Advanced Interdisciplinary Studies, Peking University, Beijing, 90871, P. R. China}
\affiliation{AI for Science Institute, Beijing 100080, P. R. China}
\author{Wenfei Li*}
\affiliation{AI for Science Institute, Beijing 100080, P. R. China}
\date{\today}

\begin{abstract}
Kohn-Sham density functional theory (DFT) is nowadays widely used for electronic structure theory simulations, and the accuracy and efficiency of DFT rely on approximations of the exchange-correlation functional. By inclusion of the kinetic energy density $\tau$, the meta-generalized-gradient approximation (meta-GGA) family of functionals achieves better accuracy and flexibility while retaining the efficiency of semi-local functionals. The SCAN meta-GGA functional has been proven to yield accurate results for solid and molecular systems.
We implement meta-GGA functionals with both numerical atomic orbitals and plane wave basis in the ABACUS package. Apart from the exchange-correlation potential, we also discuss the evaluation of force and stress. To validate our implementation, we perform finite-difference tests and convergence tests with the SCAN meta-GGA functional. We further test water hexamers, weakly interacting molecules of the S22 dataset, as well as 13 semiconductors. The results show satisfactory agreements with previous calculations and available experimental values.
\end{abstract}
\maketitle

\section{Introduction}\label{sec1}

Kohn-Sham density functional theory (DFT)\cite{64PR-Hohenberg, 65PR-Kohn} is nowadays one of the most popular paradigms in electronic structure theory. In Kohn-Sham DFT, the many-electron system is replaced by an auxiliary system of non-interacting electrons, and all many-body interactions are carried by the exchange-correlation functional $E_{xc}$. The electronic density is solved self-consistently in an iterative way, the process of which is called the self-consistent field (SCF) method. However, the exact form of the exchange-correlation functional remains unknown, and approximations have to be made. In practice, the accuracy of Kohn-Sham density functional theory simulations largely depends on the choice of the approximated exchange-correlation functional.\cite{11JCPC-Albert}

Most approximations of the exchange-correlation (XC) functionals can be classified according to the Jacob's Ladder\cite{05JCP-Perdew}. Moving up on the ladder, each rung introduces additional ingredients into the functional, making it more complicated and correcting some deficiencies of the lower rungs\cite{16JCP-Womack}, until the heaven of chemical accuracy is reached. The first three rungs of the ladder are the local density approximation (LDA), the generalized gradient approximation (GGA), and the meta-generalized gradient approximation (meta-GGA), respectively. They are classified as local or semi-local functionals because the energy functional takes the form of
$E_{\mathrm{xc}}[\rho(\textbf{r})] = \int e_{\mathrm{xc}}(\textbf{r}) d\textbf{r},$
where $\rho(\textbf{r})$ is the electron density and the exchange-correlation energy density $e_{\mathrm{xc}}(\textbf{r})$ depends only on local physical quantities at each real-space point $\textbf{r}$. As a result of this locality, the first three rungs of functionals are highly efficient in calculations by using supercomputers. The highest rung of semi-local functionals is the meta-GGA functional.\cite{13JCTC-Hao} In meta-GGA, the exchange-correlation potential depends on the kinetic energy density $\tau(\textbf{r})$ in addition to the electron density $\rho(\textbf{r})$ and its derivative $\nabla\rho(\textbf{r})$.

The inclusion of $\tau(\textbf{r})$ not only grants the meta-GGA functional to be more flexible compared to the first two rungs but also allows it to differentiate single-orbital regions from overlap regions.\cite{2003APS-Tao,2011APS-Sun,13JCTC-Hao, 13L-Sun} As a result, meta-GGA functionals are capable of producing accurate results on a variety of systems, including both molecules and solids\cite{13JCTC-Hao, 2011APS-Sun}.
%
One of the most widely used meta-GGA functionals is the SCAN functional proposed by Sun, Ruzsinszky, and Perdew.\cite{15L-Sun} The SCAN functional satisfies all 17 known physical constraints of the exact exchange-correlation functional. It has been shown to produce accurate predictions of bulk properties for semiconducting oxide, including formation enthalpies\cite{18NPJ-Zhang}, equilibrium lattice constants, cohesive energies, bulk moduli\cite{18NJP-Zhang,22PRM-Kingsbury}, and transition pressures\cite{18PRB-Shahi}. SCAN also yields qualitatively correct descriptions of liquid water and ice\cite{16NC-Sun,17NAS-Chen,19JCP-Lacount, 22JCP-Liu, 23PCCP-Liu, 22NC-Zhang}. The success of SCAN has been attributed to its ability to give a first-principles description of the medium-ranged Van der Waals interactions\cite{17NAS-Chen,19PRB-Yang}, which are absent in LDA and GGA functionals. Furthermore, when combined with the Fock exchange, the SCAN0 functional can be obtained. Recently, the SCAN0 hybrid functional has been shown to produce satisfactory results for both model systems\cite{kanungo2021comparison} and liquid water\cite{zhang2021modeling} systems.

Previously, meta-GGA functionals including the kinetic energy density have been implemented with the plane-wave (PW) basis,\cite{11B-Sun, 17JCP-Yao} and Gaussian-type orbital basis\cite{96MP-Neumann, 03L-Tao}.
Yet as far as we noticed,
an implementation of the meta-GGA functional with numerical atomic orbital (NAO) basis including forces and stress is still absent in the literature.
Compared to the widely used PW basis in condensed systems with periodic boundary conditions, 
the NAO basis\cite{10JPCM-Chen, 11JPCM-Chen, 01PRB-Anglada, 01PRB-Junquera, 10IRPC-Shang} offers highly efficient basis sets for large-scale first-principles calculations, as the number of basis functions needed is typically much smaller than the plane wave basis set.\cite{02JPCM-Josem, 09CPC-Blum, 03B-Ozaki, 08JCP-Meng}
NAOs are also strictly localized in real space, resulting in a sparse Hamiltonian matrix suitable for linear-scaling methods.\cite{99RMP-Stefan, 12RPP-Bowler}
However, complications arise when evaluating the forces and stress based on the NAO basis. 
In particular,
besides the Hellmann-Feynman terms, there are also contributions from the Pulay\cite{69MP-Pulay} and orthogonal\cite{02JPCM-Josem} terms.

In this work, we implement the meta-GGA SCAN functional for both PW and NAO basis sets in the electronic structure software ABACUS.~\footnote{https://github.com/deepmodeling/abacus-develop}
In particular, we derive the formulas of the energy term, the potential term, the forces, and the stress of the meta-GGA functionals with the systematically improvable NAO basis sets.\cite{10JPCM-Chen,11JPCM-Chen,16CMS-Li}
We follow the treatment in previous literature\cite{11B-Sun, 17JCP-Yao, 96MP-Neumann, 03L-Tao, arbuznikov2003self} in evaluating the action of meta-GGA potential on the electronic wave functions. We also note that our implementation is slightly different from that of the NAO-based ONETEP \cite{05JCP-Chris} software in the detailed implementation of meta-GGA potential term\cite{16JCP-Womack}, on which more will be elaborated in Sec.~\ref{sec2}.
We also perform systematic tests on the SCAN functional over a variety of systems. The accuracy of stress calculation is verified by comparing it with the finite-difference results.
For the plane-wave basis, we follow the formulation of Yao et al\cite{17JCP-Yao}. 

The rest of the paper is organized as follows. In Sec.~\ref{sec2}, we present the formulas of the meta-GGA and its implementation with NAO basis sets, including the corresponding numerical operations in ABACUS. Sec.~\ref{sec3} shows the results of applying the SCAN meta-GGA functional on a variety of systems. Conclusions are drawn in Sec.~\ref{sec4}. 

\section{Methods}\label{sec2}
\subsection{Evaluation of $\tau$-dependent XC Potential}
The total energy of a system, as described by the Kohn-Sham DFT\cite{65PR-Kohn}, is partitioned as
\begin{equation}
E_{\mathrm{KS}}[\rho] = T_s[\rho]+E_H[\rho]+E_{\mathrm{xc}}[\rho]+\int \rho(\textbf{r})v_{ext}(\textbf{r}) d\textbf{r},
\end{equation}
where $T_s[\rho]$, $E_H[\rho]$ and $E_{\mathrm{xc}}[\rho]$ are the non-interacting kinetic term, the Hartree term, and the exchange-correlation term, respectively. The $v_{ext}(\mathbf{r})$ term represents the external potential.
As mentioned in Sec.~\ref{sec1}, one of the key challenges in Kohn-Sham DFT is to develop an accurate and computationally efficient approximation to the exchange-correlation functional $E_{xc}$. Compared to the LDA and GGA functionals, the meta-GGA functionals are more flexible due to the inclusion of the kinetic energy density dependence.
%
The general form of $\tau$-dependent meta-GGA exchange-correlation functional is\cite{15L-Sun, 03L-Tao, 09L-Perdew, 19JCP-Bartok}
\begin{equation}
E_{\mathrm{xc}}[\rho]=\int e_{\mathrm{xc}}\Big[\rho(\mathbf{r}),\nabla\rho(\mathbf{r}),\tau(\mathbf{r})\Big]d\mathbf{r},
\end{equation}
where $\rho(\mathbf{r})$ and $\nabla\rho(\mathbf{r})$ are the electron density and its gradient, respectively, and $\tau(\mathbf{r})$ is the kinetic energy density, defined as a summation over all occupied Kohn-Sham orbitals
\begin{equation}
\tau(\mathbf{r})=\frac{1}{2}\sum_{i=1}^{occ}\left|\nabla\psi_{i}(\mathbf{r})\right|^2.
\end{equation}
To perform self-consistent electronic iterations, we need to evaluate the exchange-correlation potential $v_{\mathrm{xc}}(\mathbf{r})$ by taking the derivative of $E_{\mathrm{xc}}$ with respect to the electron density as
\begin{align}
v_{\mathrm{xc}}(\mathbf{r}) =& \frac{\delta E_{\mathrm{xc}}}{\delta\rho(\mathbf{r})} \nonumber \\
 =& \Big[\frac{\partial e_{\mathrm{xc}}}{\partial\rho}-\nabla\cdot(\frac{\partial e_{\mathrm{xc}}}{\partial\nabla\rho})\Big]+\int \frac{\delta e_{\mathrm{xc}}}{\delta\tau(\mathbf{r}')}\frac{\delta\tau(\mathbf{r'})}{\delta\rho(\mathbf{r})} d\mathbf{r'}.
 \label{eq:1}
\end{align}

In our implementation, derivatives of the energy functional,
namely $\frac{\partial e_{\mathrm{xc}}}{\partial\rho}$, $\frac{\partial e_{\mathrm{xc}}}{\partial\nabla\rho}$
and $\frac{\delta e_{\mathrm{xc}}}{\delta\tau}$, are obtained from the LIBXC package.\cite{18SX-Lehtola} Since the first two terms of Eq.~\ref{eq:1} in the square bracket also appear in GGA functionals, we focus on the last term, which is specific to the meta-GGA functionals.
We henceforth refer to it as $V^{\tau}(\mathbf{r})$, and denote the corresponding operator as $\hat{V}^{\tau}$.

Complications arise for meta-GGA functionals because the kinetic energy density $\tau$ introduces orbital dependence into the potential $V^{\tau}(\mathbf{r})$.
Although $V^{\tau}(\mathbf{r})$ is still recognized as the density functional, as kinetic energy density $\tau$ implicitly relies on the electronic density,
it is challenging to explicitly write $V^{\tau}(\mathbf{r})$ in a $\rho$-dependent form.
To be more specific, we first note that $\tau$ can be rewritten in terms of the Kohn-Sham orbitals $\{\psi_i\}$ and eigenvalues $\{\epsilon_i\}$
\begin{equation}    
\tau(\mathbf{r})=\tau[\{\psi_i\},\{\epsilon_i\}](\mathbf{r})=\frac{1}{2}\sum_{i}\Theta(\mu-\epsilon_i)|\nabla\psi_{i}(\mathbf{r})|^2, 
\end{equation}
where $\Theta$ is the Heaviside step function, and $\mu$ is the chemical potential. Then, we apply the chain rule to Eq.~\ref{eq:1} and obtain
\begin{align}
V^{\tau}(\mathbf{r})=&\int\frac{\delta e_{\mathrm{xc}}(\mathbf{r})}{\delta\tau(\mathbf{r}')}\frac{\delta\tau(\mathbf{r}')}{\delta\rho(\mathbf{r})}\mathrm{d}\mathbf{r}' \nonumber\\
=& \sum_{i=1}^{occ}\int\frac{\delta e_{\mathrm{xc}}(\mathbf{r})}{\delta\tau(\mathbf{r}')}\Big[\frac{\delta\tau(\mathbf{r'})}{\delta\psi_{i}(\mathbf{r''})}\frac{\delta\psi_{i}(\mathbf{r''})}{\delta\rho(\mathbf{r})} + \frac{\delta\tau(\mathbf{r'})}{\delta\epsilon_{i}}\frac{\delta\epsilon_{i}}{\delta\rho(\mathbf{r})} \Big]
\mathrm{d}\mathbf{r}'.
\end{align}

Therefore, in order to express $V^{\tau}(\mathbf{r})$ as a local and multiplicative potential, it is necessary to evaluate $\frac{\delta\psi_{i}(\mathbf{r'})}{\delta\rho(\mathbf{r})}$ and $\frac{\delta\epsilon_{i}}{\delta\rho(\mathbf{r})}$, which is non-trivial. One way to deal with the issue is to use the optimized effective potential (OEP) method\cite{13JCP-Zahariev}, 
which removes the explicit orbital dependence of exchange-correlation potentials by further applying the chain rule and evaluates newly derived terms using the response theory.
%
%
Although the OEP method has been applied to meta-GGA functionals \cite{yang2016more,arbuznikov2003self}, it is numerically challenging because an additional set of self-consistent equations must be solved along with the original Kohn-Sham SCF equations\cite{13JCP-Zahariev}.

Alternatively, one may observe that it is not the operator $\hat{V}^{\tau}$ itself, but rather its action on the Kohn-Sham orbitals, namely $\hat{V}^{\tau}\psi_{i}$, that is of practical importance. Therefore, by inserting
\begin{equation}
    \frac{\delta\rho(\mathbf{r})}{\delta\psi_{i}(\mathbf{r'})}=2\psi_{i}(\mathbf{r'})\delta(\mathbf{r}-\mathbf{r'}),
\end{equation}
into $\hat{V}^{\tau}\psi_{i}$, one arrives at
\begin{eqnarray}
\hat{V}^{\tau}\psi_{i}(\mathbf{r}) & = & \frac{1}{2}\int\frac{\delta e_{\mathrm{xc}}}{\delta\tau(\mathbf{r'})}\frac{\delta\tau(\mathbf{r'})}{\delta\psi_{i}(\mathbf{r})}\mathrm{d}\mathbf{r'}\nonumber \\
 & = & -\frac{1}{2}\nabla\cdot\left[\frac{\delta e_{\mathrm{xc}}}{\delta\tau(\mathbf{r})}\nabla\psi_{i}(\mathbf{r})\right].\label{eq:2}
\end{eqnarray}
By doing so, we convert the original problem of evaluating functional derivative of $\frac{\delta e_{xc}}{\delta \rho}$ into evaluating $\frac{\delta e_{xc}}{\delta \psi}$, 
which could be explicitly written out.
However, this operation turns $V^{\tau}(\mathbf{r})$ into a non-multiplicative operator
\begin{equation}
    \hat{V}^{\tau}(\mathbf{r})=-\frac{1}{2}\nabla\cdot\left[\frac{\delta e_{xc}}{\delta \tau(r)}\nabla\right],
\end{equation}
which implies that the value of $\tau$-dependent XC potential varies depending on the Kohn-Sham wave function $\psi_i$.\cite{03CPL-Alexei}
The formulation in Eq.~\ref{eq:2} was first brought up in Ref.~\onlinecite{96MP-Neumann} to evaluate the Becke-Roussel exchange functional\cite{89A-Becke} and has been widely adopted in the implementations of meta-GGA functionals.\cite{89A-Becke, 03L-Tao, 98JCP-Troy, 02PCCP-Alexei, 11B-Sun, 17JCP-Yao} It has been given different names in the literature, such
as the "orbital-based density-functional derivative method (ODDM)", \cite{13JCP-Zahariev} or "functional derivatives of $\tau$-dependent functionals with respect to the orbitals (FDO)". \cite{16JCP-Womack, 03CPL-Alexei} In this work, we adopt the same framework.

\subsection{Kinetic energy density and Hamiltonian in NAOs}

In numerical atomic basis, the Kohn-Sham orbitals are expanded as a linear combination
of atomic orbitals as
\begin{equation}
\psi_{i}(\mathbf{r})=\sum_{\mu}C_{i\mu}\chi_{\mu}(\mathbf{r}),
\end{equation}
where $C_{i\mu}$ denotes the coefficients, $\chi_\mu$ depicts atomic orbitals, and $i$ is the index for the Kohn-Sham wave function $\psi_{i}(\mathbf{r})$.
Each atomic basis is located on a certain atom in the system: $\chi_{\mu}(\mathbf{r})=\bar{\chi}_{\mu}(\mathbf{r}-\mathbf{X_{a}})$,
where $\mathbf{X_{a}}$ is the position of atom $a$.
In ABACUS, systems are treated with periodic boundary conditions (PBCs) with multiple $k$-point sampling. For clarity of discussion, we start with a derivation
in the non-periodic case, followed by a short extension to the periodic scenario.

We discuss the two major added components of the meta-GGA functional as compared to the GGA functionals, i.e., the kinetic
energy density $\tau$ and the contribution of the $\tau$-dependent
part to the Hamiltonian matrix $\langle\chi_{\mu}|\hat{V}_{\tau}|\chi_{\nu}\rangle$.

First, the $\tau$ term is given by
\begin{align}
\tau(\mathbf{r}) & =\frac{1}{2}\sum_{i=1}^{occ}f_i[\nabla\psi_{i}(\mathbf{r})]^{*}\cdot[\nabla\psi_{i}(\mathbf{r})]\nonumber\\
 & =\frac{1}{2}\sum_{\mu\nu}\rho_{\mu\nu}[\nabla\chi_{\mu}(\mathbf{r})]\cdot[\nabla\chi_{\nu}(\mathbf{r})]\label{kinetic-density},
\end{align}
where $f_{i}$ is the occupation number of state $i$ while the density matrix $\rho_{\mu\nu}$ takes the form of 
\begin{equation}
\rho_{\mu\nu}=\sum_{i=1}^{occ}f_{i}C_{i\mu}^{*}C_{i\nu}.
\end{equation}
%
%
Under the PBCs, Kohn-Sham wave functions obey the Bloch theorem and take the form of
\begin{equation}
\psi_{i\mathbf{k}}(\mathbf{r}) = \frac{1}{\sqrt{N}}\sum_{\mathbf{R}}\sum_{\mu}C_{i\mu}(\mathbf{k})\chi_{\mu \mathbf{R}}(\mathbf{r})e^{i\mathbf{k}\cdot\mathbf{R}},
\end{equation}
where $N$ is the number of unit cells in the Born-von-Karman supercell, $\chi_{\mu\mathbf{R}}=\bar{\chi}_{\mu}(\mathbf{r}-\mathbf{X}_a-\mathbf{R})$ is the atomic orbital located on atom $a$ in the cell $\mathbf{R}$, and $\mathbf{k}$ labels the $k$-points in the first Brillouin zone.
With this, the density matrix in real space takes the form of
\begin{equation}
    \rho_{\mu\nu}(\mathbf{R})=\frac{1}{N_{\mathbf{k}}}\sum_{\mathbf{k}}\sum_{i=1}^{occ}f_{i}C_{i\mu}^{*}(\mathbf{R})C_{i\nu}(\mathbf{R})e^{-i\mathbf{k}\cdot\mathbf{R}},
    \label{rho_pbc}
\end{equation}
where $N_{\mathbf{k}}$ denotes the number of $k$-points sampled in the first Brillouin zone.
In this regard, the kinetic energy density in PBCs is written as
\begin{align}
    \tau(\mathbf{r})&=\frac{1}{2}\frac{1}{N_{\mathbf{k}}}\sum_{\mathbf{k}}\sum_{i=1}^{occ}f_{i}[\nabla\psi_{i\mathbf{k}}(\mathbf{r})]^{*}\cdot[\nabla\psi_{i\mathbf{k}}(\mathbf{r})]\\
    &= \frac{1}{2}\sum_{\mu\nu}\sum_{\mathbf{R}}\rho_{\mu\nu}(\mathbf{R})[\nabla\chi_{\mu\mathbf{0}}(\mathbf{r})]\cdot[\nabla\chi_{\nu\mathbf{R}}(\mathbf{r})].
\end{align}    

Second, the $\tau$-dependent XC matrix element,  $\langle\chi_{\mu}|\hat{V}_{\tau}|\chi_{\nu}\rangle$ term, 
is obtained
using integration by parts
\begin{align}
\hat{V}_{\mu\nu}^{\tau} & =\langle\chi_{\mu}|\hat{V}^{\tau}|\chi_{\nu}\rangle\nonumber \\
 & =-\frac{1}{2}\int \chi_{\mu}(\mathbf{r})\nabla\cdot \left[\frac{\delta e_{\mathrm{xc}}}{\delta\tau(\mathbf{r})}\nabla\chi_{\nu}(\mathbf{r})\right]\mathrm{d}\mathbf{r}\label{eq:3}\\
 & =\frac{1}{2}\int [\nabla\chi_{\mu}(\mathbf{r})]\cdot[\nabla\chi_{\nu}(\mathbf{r})]\frac{\delta e_{\mathrm{xc}}}{\delta\tau(\mathbf{r})}\mathrm{d}\mathbf{r}.\label{eq:3_1} 
\end{align}
We note that such derivation is first brought up by Ref.~\onlinecite{96MP-Neumann}. Eq.~\ref{eq:3_1} has been implemented in the Gaussian software\cite{96MP-Neumann, 98JCP-Troy}, and the derivative of Gaussian orbital could be derived analytically.
Meanwhile, the ONETEP team implemented the formula of Eq.~\ref{eq:3}, where the gradient operator has been applied to the reciprocal space due to the derivative of the non-orthogonal generalized Wannier function basis adopted in ONETEP may lose its locality.\cite{16JCP-Womack} 
With PBCs, the $\tau$-dependent contribution in real space is written as 
\begin{equation}
    \hat{V}_{\mu\nu}^{\tau}(\mathbf{R}) =\frac{1}{2}\int [\nabla\chi_{\mu\mathbf{0}}(\mathbf{r})]\cdot[\nabla\chi_{\nu\mathbf{R}}(\mathbf{r})]\frac{\delta e_{\mathrm{xc}}}{\delta\tau(\mathbf{r})}\mathrm{d}\mathbf{r}.
\end{equation}
When obtaining the expansion coefficients of Kohn-Sham orbitals $C_{i\mu}(\mathbf{k})$,
the operator is transformed into the corresponding $k$-space
representation $\hat{V}_{\mu\nu}^{\tau}(\mathbf{k})=\sum_{\mathbf{R}}\hat{V}_{\mu\nu}^{\tau}(\mathbf{R})e^{-i\mathbf{k}\cdot\mathbf{R}}$ through Fourier Transform.
%

In ABACUS, derivatives are directly evaluated from the numerical atomic orbitals, so we adopt Eq.~\ref{eq:3_1}.
Additionally, we note that the evaluation of both $\tau(\mathbf{r})$ and $\hat{V}_{\mu\nu}^{\tau}$ require expanding the basis functions
$\{|\chi_{\mu}\rangle\}$ (and their derivatives) in real space. 
ABACUS achieves operations involving such expansions with real-space FFT (Fast Fourier Transform) grid points as the basic operation unit. 
A more detailed discussion of such operations can be found in Section~\ref{grid_int}. 

With the above details of evaluating $\tau$ and $\hat{V}_{\tau}$, we are able to run SCF calculations using meta-GGA functionals. However, to enable structural relaxation or {\it ab-initio} molecular dynamics, the forces and stresses are needed.

\subsection{$\tau$-dependent Force and Stress in NAOs}
The $\tau$-dependent force term of atom $a$ takes the form of
\begin{align}
F_{a}^{\tau} & =-\frac{\mathrm{d}}{\mathrm{d}\mathbf{X}_{\mathbf{a}}}\mathrm{Tr}[\hat{V}^{\tau}]\nonumber\\
 & =-\frac{\mathrm{d}}{\mathrm{d}\mathbf{X_{a}}}\sum_{\mu\nu}\rho_{\mu\nu}\langle\chi_{\mu}|\hat{V}^{\tau}|\chi_{\nu}\rangle.
\end{align}
In practice, the derivative operation results in three terms. The first term is the derivative of the density matrix $\rho_{\mu\nu}$ with respect to the atom positions, which is also known as the "orthogonal" term\cite{02JPCM-Josem}. The term vanishes if an orthogonal set of basis functions is used.
The second term is the derivative of the operator with respect to the atom positions, namely the Hellmann-Feynman term. The last one is the Pulay term\cite{69MP-Pulay}, involving the derivative of basis functions with respect to atom positions.

In fact, since the $V^{\tau}(\mathbf{r})$ term is independent of the atom positions, the Hellmann-Feynman term vanishes. In particular, when the plane-wave basis is employed, the Pulay and orthogonal terms are absent, so the $\tau$-dependent part of meta-GGA functionals has no contributions to the total force.

For NAO basis sets, the Pulay and orthogonal terms need to be considered. As discussed in the literature\cite{02JPCM-Josem}, the orthogonal term arises due to the requirement for maintaining the orthogonality of Kohn-Sham orbitals $\{\psi_{i}\}$, and it is obtained using the following transformation
\begin{equation}
\sum_{\mu\nu}\hat{H}_{\mu\nu}\frac{\mathrm{d}}{\mathrm{d}\mathbf{X}_{\mathbf{a}}}\rho_{\mu\nu}=-\sum_{\mu\nu}E_{\mu\nu}\frac{\mathrm{d}}{\mathrm{d}\mathbf{X}_{\mathbf{a}}}S_{\mu\nu},
\end{equation}
where $S_{\mu\nu}=\langle\chi_{\mu}|\chi_{\nu}\rangle$ is the overlap matrix of atomic basis. The orthogonal term is evaluated for all components of the Hamiltonian matrix, and no extra work is required specifically for the $\tau$-dependent part of meta-GGA functionals.

The Pulay term is given by
\begin{align}
F_{a}^{\tau,Pulay} & =-\sum_{\mu\nu}\rho_{\mu\nu}\Big[\left\langle\frac{\mathrm{d}}{\mathrm{d}\mathbf{X_{a}}}\chi_{\mu}\Big|\hat{V}^{\tau}\Big|\chi_{\nu}\right\rangle\nonumber \\
&+\left\langle\chi_{\mu}\Big|\hat{V}^{\tau}\Big|\frac{\mathrm{d}}{\mathrm{d}\mathbf{X_{a}}}\chi_{\nu}\right\rangle\Big]\nonumber\\
 & =-2\sum_{\mu\nu}\rho_{\mu\nu}\left\langle\frac{\mathrm{d}}{\mathrm{d}\mathbf{X_{a}}}\chi_{\mu}\Big|\hat{V}^{\tau}\Big|\chi_{\nu}\right\rangle.\label{eq:4}
\end{align}
Here we know the density matrix $\rho_{\mu\nu}$ is symmetric. As for the derivative of basis functions with respect to the atom positions, the derivative is nonzero only for basis functions that are located on atom $a$, and for $\mu\in a$, we have
\begin{align}
\frac{\mathrm{d}}{\mathrm{d}\mathbf{X_{a}}}\chi_{\mu}(\mathbf{r}) & =\frac{\mathrm{d}}{\mathrm{d}\mathbf{X_{a}}}\bar{\chi}_{\mu}(\mathbf{r}-\mathbf{X_{a}})\nonumber \\
 & =-\nabla\bar{\chi}_{\mu}(\mathbf{r}-\mathbf{X_{a}})\nonumber \\
 & =-\nabla\chi_{\mu}(\mathbf{r}).\label{eq:5}
\end{align}
Substituting Eq. \ref{eq:3_1} and Eq. \ref{eq:5} into Eq. \ref{eq:4}, we obtain the expression
of Pulay force on atom $a$ along the $\alpha$ direction ($\alpha and \beta$ take the values of $x,y,z$)
\begin{align}
&F_{a\alpha}^{\tau,Pulay}=2\sum_{\mu\in a\nu}\rho_{\mu\nu}\langle\frac{\partial}{\partial\alpha}\chi_{\mu}\Big|\hat{V}^{\tau}\Big|\chi_{\nu}\rangle\nonumber\\
 & =\sum_{\mu\in a\nu}\rho_{\mu\nu}\int \frac{\delta e_{\mathrm{xc}}}{\delta\tau(\mathbf{r})}\sum_{\beta}\left[\frac{\partial^{2}}{\partial\alpha\partial\beta}\chi_{\mu}(\mathbf{r})\right]\left[\frac{\partial}{\partial\beta}\chi_{\nu}(\mathbf{r})\right]\mathrm{d}\mathbf{r}.
\end{align}
Similar to the evaluation of Hamiltonian element evaluation under PBCs, the Pulay force under the PBCs takes the form of
\begin{align}    
F_{a\alpha}^{\tau,Pulay}=&\sum_{\mathbf{R}}\sum_{\mu\in a\nu}\rho_{\mu\nu}(\mathbf{R})\frac{\delta e_{\mathrm{xc}}}{\delta\tau(\mathbf{r})}\nonumber\\
&\sum_{\beta}\left[\frac{\partial^{2}}{\partial\alpha\partial\beta}\chi_{\mu\mathbf{0}}(\mathbf{r})\right]\left[\frac{\partial}{\partial\beta}\chi_{\nu\mathbf{R}}(\mathbf{r})\right]\label{XC-force},
\end{align}
where $\rho_{\mu\nu}(\mathbf{R})$ is defined in Eq.~\ref{rho_pbc}.

Similarly, the $\tau$-dependent part contributes to the orthogonal and Pulay terms in the total stress. On the one hand, the orthogonal term is calculated along with other components of the Hamiltonian as noted in a previous work\cite{21CPC-Zheng}. 
On the other hand, the Pulay stress in the $\alpha\beta$ direction is ($\alpha,\beta,\gamma$ take values of $x,y,z$)
\begin{align}
\sigma_{\alpha\beta}^{\tau,Pulay}=&-\frac{1}{2\Omega}\sum_{\mu\nu}\rho_{\mu\nu}\int \frac{\delta e_{\mathrm{xc}}}{\delta\tau(\mathbf{r})}(r^{\beta}-X_{a}^{\beta})\nonumber\\
&\sum_{\gamma}\left[\frac{\partial^{2}}{\partial\alpha\partial\gamma}\chi_{\mu}(\mathbf{r})\right]\left[\frac{\partial}{\partial\gamma}\chi_{\nu}(\mathbf{r})\right]\mathrm{d}\mathbf{r},
\end{align}
where $\Omega$ is the volume of cell, while $r^{\beta}$ and $X_{a}^{\beta}$ denote the components of $\mathbf{r}$ and $\mathbf{X_{a}}$ in the $\beta$ direction.
Furthermore, by considering the PBCs, the Pulay stress can be written as
\begin{align}
    \sigma_{\alpha\beta}^{\tau,Pulay}=&-\frac{1}{2\Omega}\sum_{\mathbf{R}}\sum_{\mu\nu}\int\frac{\delta e_{\mathrm{xc}}}{\delta\tau(\mathbf{r})}(r^{\beta}-X_{a}^{\beta})\nonumber\\
&\sum_{\gamma}\left[\frac{\partial^{2}}{\partial\alpha\partial\gamma}\chi_{\mu\mathbf{0}}(\mathbf{r})\right]\left[\frac{\partial}{\partial\gamma}\chi_{\nu\mathbf{R}}(\mathbf{r})\right]\mathbf{\mathrm{d}r}.\label{XC-stress}
\end{align}
From the expressions of Pulay force and stress terms, we note that they also require the evaluation of atomic basis functions and their derivatives in real space; more details are presented in Sec.~\ref{grid_int}.

\begin{figure}[htp]
    \includegraphics[width=8.5cm]{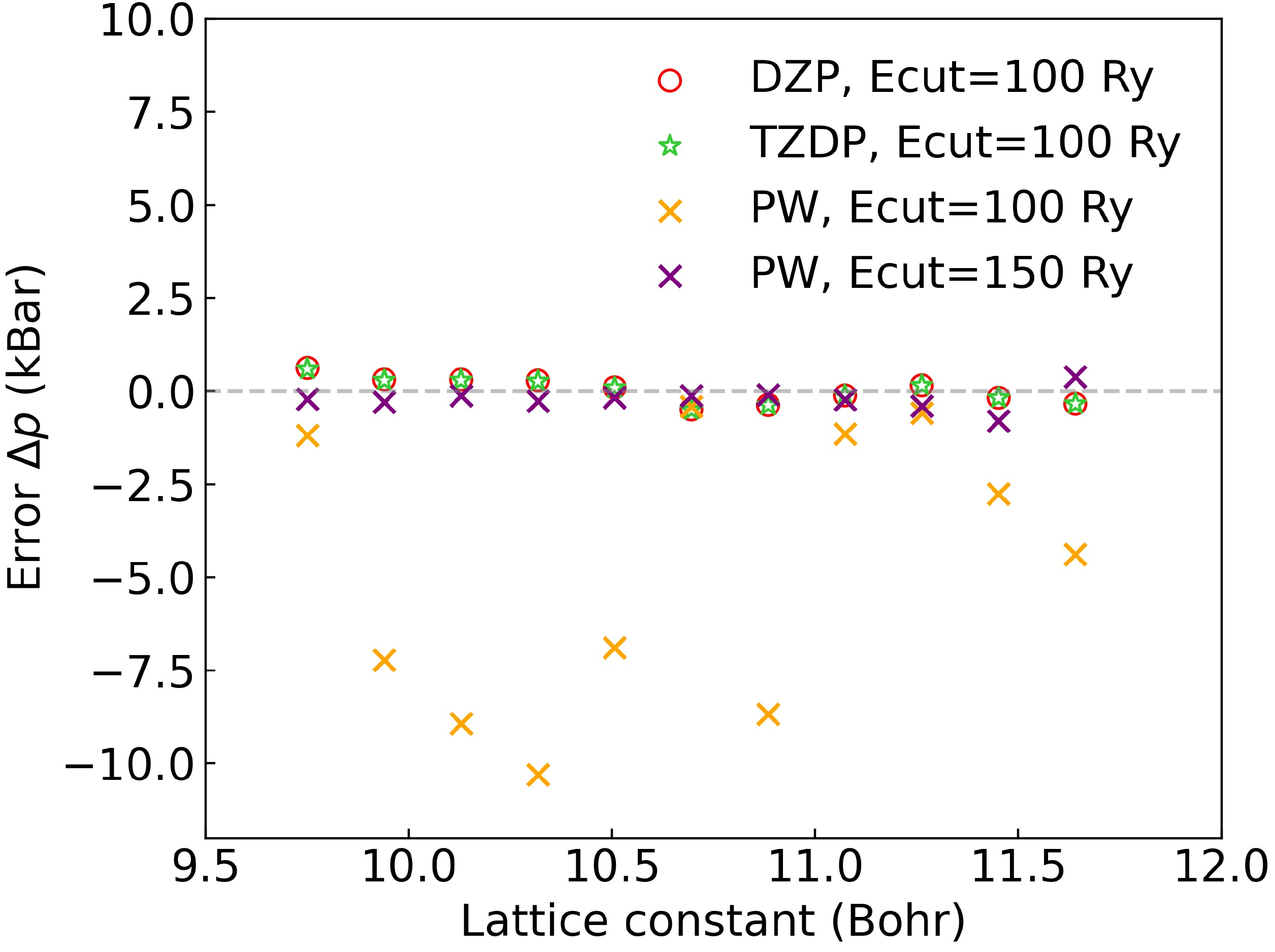}
    \caption{Stress errors $\Delta p$ (in kBar) of GaAs with respect to different lattice constants. Both NAO basis sets (DZP and TZDP) and PW basis sets are used in ABACUS. The error is defined as the difference between the stress computed by the finite-difference method (set to 0) and the stress computed by the analytic method.
    The parameter Ecut in the legend denotes the kinetic energy cutoff adopted in these calculations.}\label{stress_fd}
\end{figure}

\subsection{Operations on Real-Space Grids}\label{grid_int}
In ABACUS, in order to obtain the NAO values on real-space grids and their derivatives with respect to atom positions, we adopt uniform real-space grid points $\{\mathbf{r}\}$ under PBCs as basic units of grid operations.
Values calculated on these grids are either directly assigned with physical quantities such as the electron density $\rho(\mathbf{r})$ or the kinetic energy density $\tau(\mathbf{r})$, or accumulated (integrated) as in the case of matrix elements, force or stress.
The major physical quantities include the $\tau$-dependent XC functional, the kinetic energy density $\tau$ (Eq.~\ref{kinetic-density}), the $\tau$-dependent exchange-correlation potential $V_{\mu\nu}^{\tau}$ (Eq.~\ref{eq:3_1}), the $\tau$-dependent forces $F_{a\alpha}^{\tau,Pulay}$ (Eq.~\ref{XC-force}), and the $\tau$-dependent stress $\sigma_{\alpha\beta}^{\tau,Pulay}$ (Eq.~\ref{XC-stress}).
These physical quantities can be evaluated based on three classes of basic grid operations:
evaluation of basis orbitals $\{\chi_\mu(\textbf{r})\}$ and their derivatives, transforming a vector by the density matrix: $f_{\mu}(\mathbf{r})=\sum_{\nu}\rho_{\mu\nu}g_{\nu}(\mathbf{r})$, and arithmetic operations.

For the first operation, i.e., evaluation of $\chi_{\mu}$ and its derivatives on the grid points, we separate the basis function into the radial and the angular parts,
\begin{equation}
\chi_{\mu}(\mathbf{r})=R_{\mu}(r)Y_{lm}(\theta,\phi),
\end{equation}
where the radial part depicts a linear combination of spherical Bessel functions generated by minimizing the spillage of the wave functions between the atomic orbital calculations and the converged plane wave calculations for dimer systems~\cite {10JPCM-Chen}. Furthermore, the angular part denotes the real spherical harmonics. 

As for the derivatives of numerical atomic orbitals,
we first make the following transformation,
\begin{equation}
\chi_{\mu}(\mathbf{r})=\frac{R_{\mu}(r)}{r^{l}}\cdot \left[r^{l}Y_{lm}(\theta,\phi)\right].
\end{equation}
\begin{figure}
    \includegraphics[width=8.5cm]{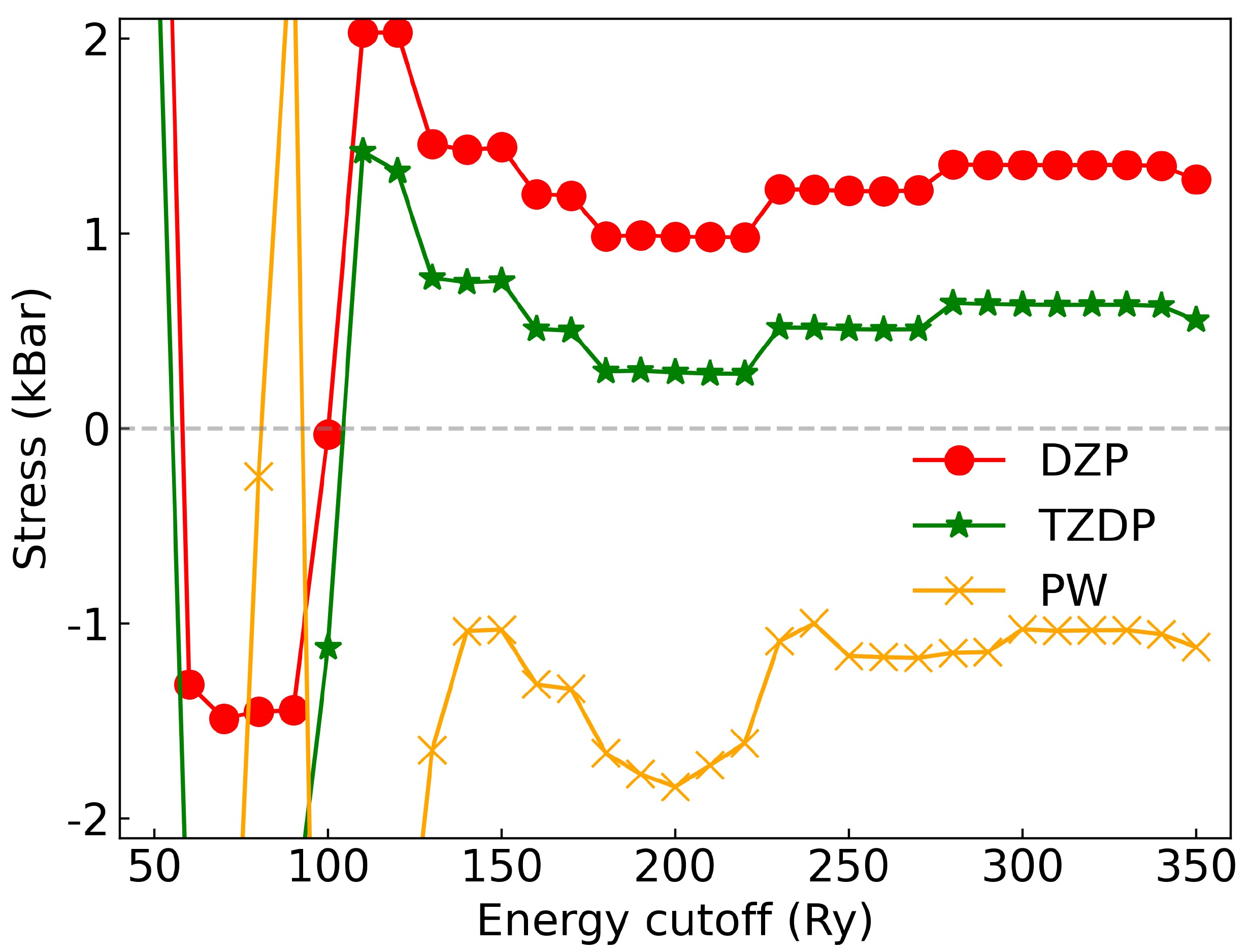}
    \caption{Stresses of GaAs calculated with the PW (the orange line), DZP (the red line), and TZDP (the green line) basis sets with different energy cutoffs for the uniform real-space grid.}\label{ecut_converg}
\end{figure}
The term in the square bracket gives the so-called `solid' spherical harmonics, which can be readily expressed in the Cartesian coordinates. For example, the solid spherical harmonics with $l=2,m=2$ can be written as
\begin{equation}
r^{2}Y_{22}(\theta,\phi)=\frac{1}{4}\sqrt{\frac{15}{\pi}}(x^{2}-y^{2}),
\end{equation}
and the gradient is 
\begin{equation}
    \nabla[r^{2}Y_{22}]=\frac{1}{4}\sqrt{\frac{15}{\pi}}(2x\hat{x}-2y\hat{y}).
\end{equation}
Furthermore, the gradients of the basis functions are calculated using the product rule
\begin{align}
\nabla\chi_{\mu} & =\left(\nabla\frac{R_{\mu}}{r^{l}}\right)\cdot r^{l}Y_{lm}+\frac{R_{\mu}}{r^{l}}\cdot \nabla(r^{l}Y_{lm})\nonumber\\
 & =\hat{r}\frac{R_{\mu}'-lR_{\mu}}{r^{l+1}}\cdot(r^{l}Y_{lm})+\frac{R_{\mu}}{r^{l}}\cdot\nabla(r^{l}Y_{lm}).
\end{align}

As for the second operation, i.e., $f_{\mu}(\mathbf{r})=\sum_{\nu}\rho_{\mu\nu}g_{\nu}(\mathbf{r})$ (transforming a vector by the density matrix), we note that in ABACUS, the density matrix is stored in a form, so the basis functions on a pair of atoms $(a_{1}, a_{2})$ define a contiguous sub-matrix. Therefore, we partition the vector $g_{\nu}(\mathbf{r})$ into segments according to the atoms to which the index $\nu$ belongs. Matrix-vector multiplications are then carried out by calling the LAPACK\cite{99-Anderson} subroutines.
%

With the above operations, we are able to calculate needed physical quantities.
Take the Pulay force in Eq.~(\ref{XC-force}) as an example,
we perform the following five steps for each grid point to evaluate the Pulay force.
First, we evaluate the derivatives terms $\frac{\partial^{2}}{\partial\alpha\partial\beta}\chi_{\mu}(\mathbf{r})$ and $\frac{\partial}{\partial\beta}\chi_{\nu}(\mathbf{r})$.
Second, a multiplication operation is performed to obtain $g_{\nu\beta} (\textbf{r}) = \frac{\delta e_{\mathrm{xc}}}{\delta\tau(\mathbf{r})}\frac{\partial}{\partial\beta}\chi_{\nu}(\mathbf{r})$.
Third, we transform $g_{\nu\beta}$ by the use of the density matrix: $f_{\mu\beta} = \sum_{\nu}\rho_{\mu\nu}g_{\nu\beta}$.
Fourth, another multiplication operation is performed: $h_{\alpha\beta} = \sum_{\mu}\frac{\partial^{2}\chi_{\mu}}{\partial\alpha\partial\beta} f_{\mu\beta}$.
Finally, $h_{\alpha\beta}$ is accumulated to generate the corresponding force component.

In practice, several adjacent grid points (typically a 2$\times$2$\times$2
cube) are grouped to exploit the efficiency of
matrix-matrix multiplications. Furthermore, parallelization is achieved
by distributing the independent units onto different processes.

\begin{figure}
    \includegraphics[width=8.5cm]{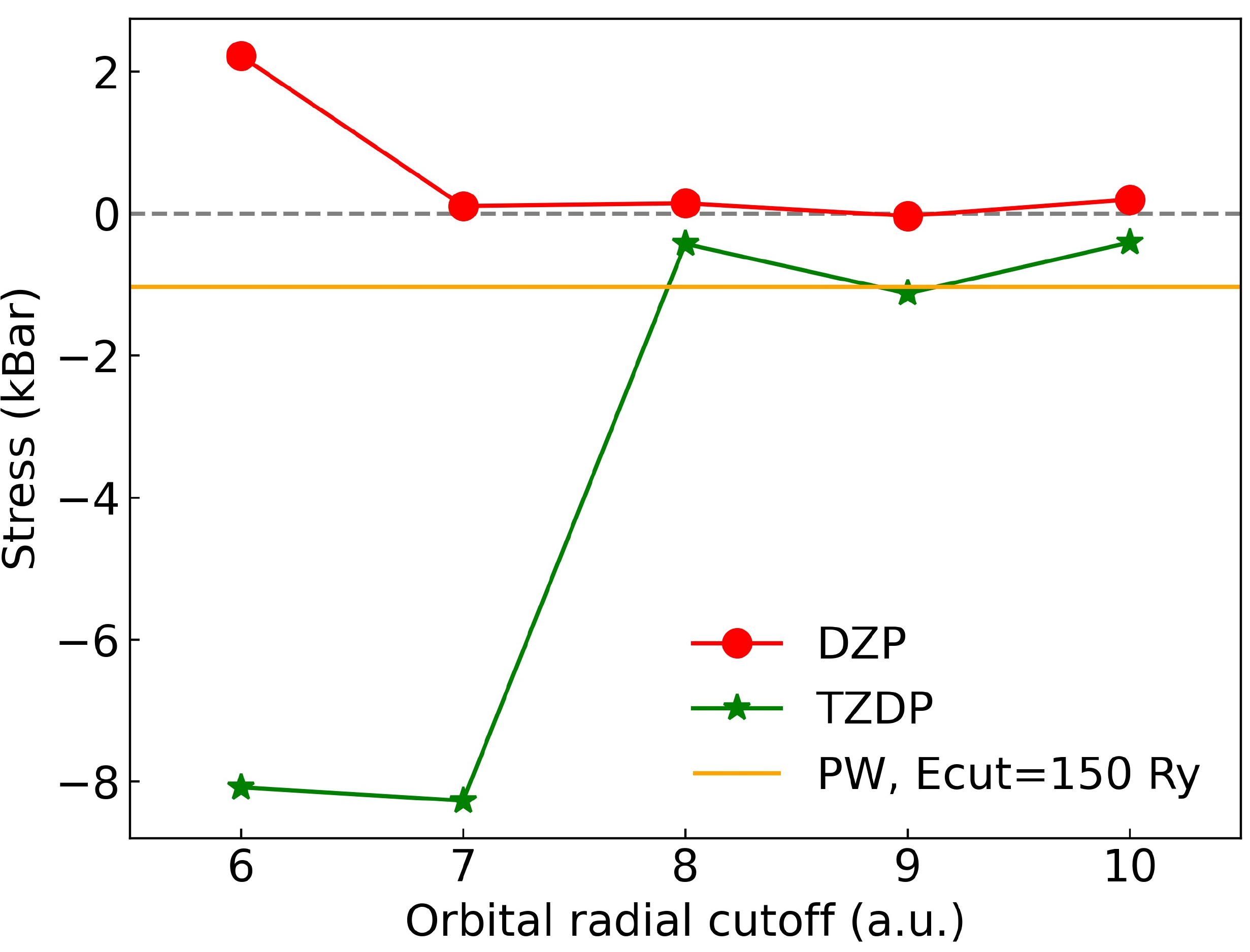}
    \caption{Stress of GaAs calculated with the DZP (red solid line) and the TZDP (green solid line) basis sets with different radius cutoffs ranging from 6.0 to 10.0 a.u. The orange line marks the stress calculated with the PW basis and 150 Ry energy cutoff (Ecut).}\label{rcut_converg}
\end{figure}

\section{Results and Discussions}\label{sec3}

Previous studies showed that the SCAN functional yields significantly more accurate results on such systems as compared to the PBE functional.\cite{15L-Sun, 16X-Peng}
Here we first compare the results from the SCAN functional with those obtained from the finite difference calculations. We also study the convergence behavior of stress with respect to the radius cutoff of numerical atomic orbitals and the energy cutoff of the FFT grid.
%
We use the ABACUS 3.0.3 package together with the LIBXC 5.2.3 package to perform meta-GGA calculations.
We validate a series of materials, including the water hexamers, the weakly interacting molecules of the S22 dataset\cite{06PCCP-Petr}, as well as 13 semiconducting materials. 

\begin{figure}[htp]
    \includegraphics[width=8.5cm]{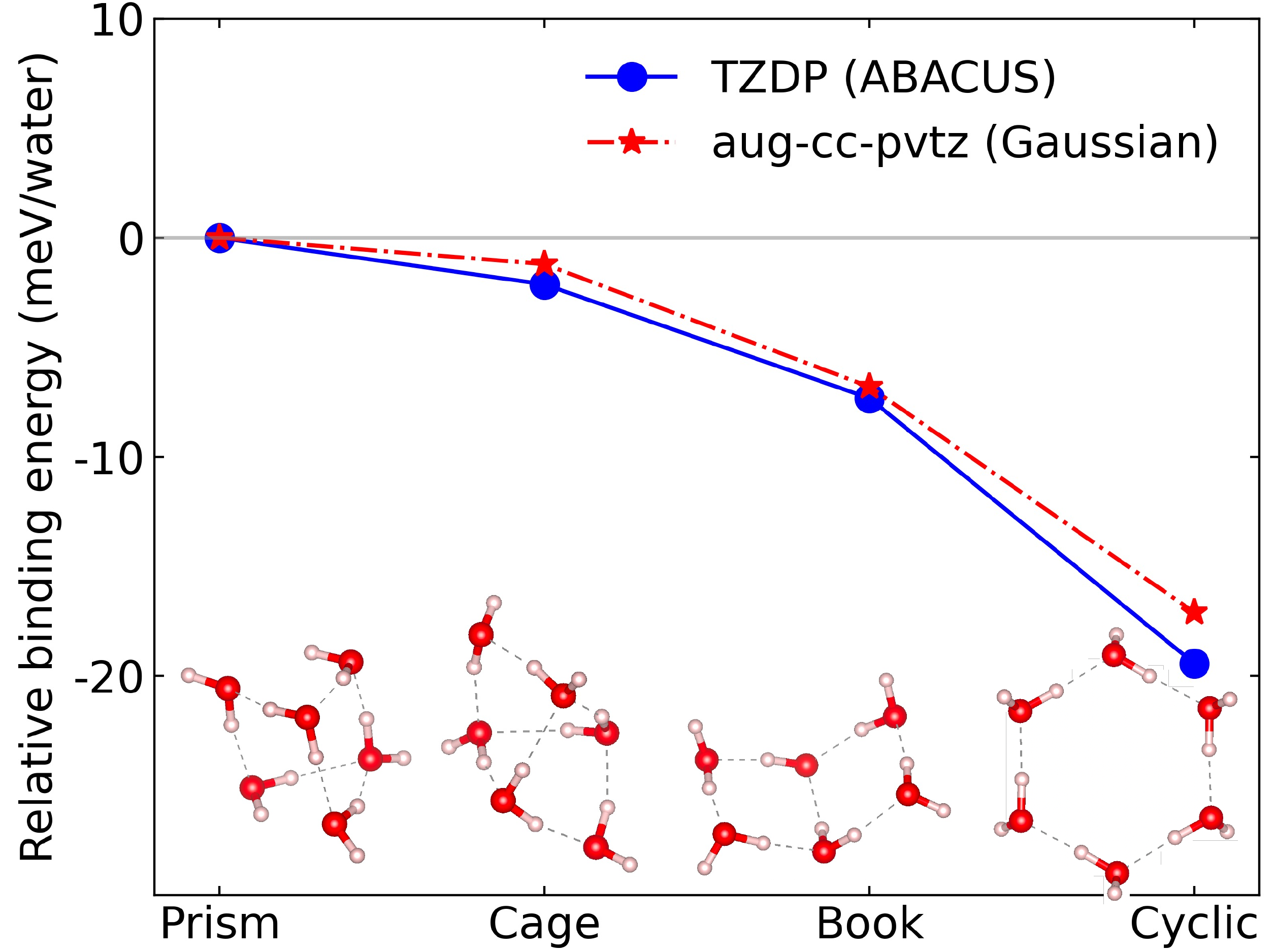}
    \caption{Relative binding energies of water hexamers with four different structures, i.e., the prism, cage, book, and cyclic. We set the binding energy of the prism structure as zero. The values calculated from the ABACUS package with the use of the SCAN XC functional and TZDP basis set are in blue. The red line is from Ref.~\onlinecite{16NC-Sun}, where the Gaussian package with the aug-cc-pvtz basis sets is adopted.}\label{water_cluster}
\end{figure}

\begin{table*}[htp]
\renewcommand{\arraystretch}{1.5}
\setlength{\tabcolsep}{4mm}
\caption{Two different types of numerical atomic orbitals for 13 elements, i.e., DZP and TZDP, used in calculations. The number of valence electrons and the radius cutoff for NAOs (in a.u.) are marked in the parenthesis for each element. In terms of the DZP and TZDP basis sets, the number of NAOs for each angular momentum is shown, and the total number of atomic orbitals for each atom is listed in the parenthesis. In semiconductor calculations, we use 6 and 7 a.u. NAOs for the C and N, respectively. In the water hexamers and S22 tests, we adopt 10 a.u. NAOs for the C and N.}
    \begin{center}
    \begin{tabular}{lcc}
    \hline
    \hline
          & DZP & TZDP \\
    \hline
    H (1e, 10 a.u.) & 2s, 3p (5 orbitals) & 3s, 6p (9 orbitals)\\
    \hline
    C (4e, 6/10 a.u.), N (5e, 7/10 a.u.), O (6e, 10 a.u.),  & 2s, 6p, 5d (13 orbitals) & 3s, 9p, 10d (22 orbitals)\\
    Si (4e, 8 a.u.), P (5e, 9 a.u.), As (5e, 9 a.u.) &  &  \\
    \hline
    Ga (13e, 9 a.u.), Ge (14e, 9 a.u.),  & 2s, 6p, 10d, 7f (25 orbitals) & 3s, 9p, 15d, 14f (41 orbitals)\\
    In (13e, 9 a.u.), Sb (15e, 9 a.u.) &  &  \\
    \hline
    Al (11e, 9 a.u.) & 4s, 12p, 5d (21 orbitals) & 6s, 18p, 10d (34 orbitals) \\   
    \hline
    \end{tabular}
    \end{center}
    \label{orbitals}
\end{table*}

In all of the tests, an energy cutoff of 100 Ry is used unless specially noted. The optimized Norm-conserving Vanderbilt (ONCV) pseudopotentials\cite{15CPC-Schlipf} generated by the Perdew-Burke-Ernzerhof\cite{96L-Perdew} (PBE) exchange-correlation functional are used throughout the tests. We adopt two sets of NAO basis sets, namely, the double zeta orbitals with a polar orbital (DZP) and the triple zeta orbitals with two polar orbitals (TZDP). A detailed number of orbitals used in the DZP and TZDP basis sets are listed in Table.~\ref{orbitals} for 13 elements.

DFT calculations on four water hexamers and weakly interacting clusters are carried out in a 20$\times$ 20$\times$20 \AA$^3$ cell,
where the $\Gamma$-point sampling of the Brillouin zone is used. 
We use a radius cutoff of 10 a.u. for the TZDP basis sets in order to achieve the desired level of accuracy in the two tests, as shown in Table~\ref{orbitals}.
In the test of semiconductors, the radius cutoffs of elements are also listed in Table.~\ref{orbitals}. 
In addition, a Monkhorst-Pack (MP) $k$-mesh of 6$\times$6$\times$6 is used for self-consistent calculations.
In the tests of the water hexamers and semiconductors, the structures are fully relaxed until the largest atomic force is less than 0.01 eV/\AA.

\subsection{Finite-Difference Test and Convergence Test of Stress}
We compare the analytic stress results with those obtained from the finite-difference (FD) method.
The stress tensor is defined as
\begin{align}
    \sigma_{\alpha\beta} &= -\frac{1}{\Omega}\frac{\partial E_{tot}}{\partial \epsilon_{\alpha\beta}}|_{\epsilon=0},
\end{align}
where $\alpha, \beta =x, y, z$. The strain tensor $\epsilon_{\alpha\beta} $ stands for an infinitesimal displacement of the crystal lattice, and $\Omega$ depicts the volume of the cell.

We take GaAs as an example and Fig.~\ref{stress_fd} shows the comparison results for the calculated stresses obtained by using both NAO and PW basis sets in ABACUS. 
The stress difference is obtained by using the analytic method and the FD method.
The pressure $p=\frac{1}{3}\sum_{\alpha=1}^{3}\sigma_{\alpha\alpha}$ is calculated by using the FD method, where we apply $\pm 0.15\%$ isotropic deformation for the cell.
We note the errors obtained from the PW basis with an energy cutoff of 100 Ry are larger than 5 kBar for several cases. This is caused by the Pulay stress arising from the variation of the plane wave basis set under strain, which is not considered in the finite difference calculations. 
Notably, the effect diminishes when we use a converged basis set.\cite{21CPC-Zheng, caro2012comparison} By increasing the energy cutoff to 150 Ry, the errors of PW calculations are reduced to the order of 0.1 kBar. On the other hand, the errors in NAO basis are much smaller at 100 Ry, already at the order of 0.1 kBar. The results are expected because in DFT calculations with NAO basis sets, the energy cutoff only affects the number of real-space grid points; since the basis functions remain the same, no additional Pulay stress is present.

The effect of Pulay stress is also reflected in the slow convergence of stress with energy cutoff for PW basis\cite{21CPC-Zheng}. As shown in Fig.~\ref{ecut_converg}, the stresses calculated with both DZP and TZDP basis converge to within 1 kBar with an energy cutoff of 120 Ry, while a cutoff of 230 Ry is required for PW calculations to achieve the same level of accuracy. Also, the convergence of stress using the SCAN functional in PW basis is slower than that of the PBE functional\cite{21CPC-Zheng}, where convergence at 1 kBar is obtained at an energy cutoff of 150 Ry. Indeed, it has been shown that the SCAN functional is numerically less stable than GGA functionals, and often requires a denser real-space grid.\cite{20JPCL-Furness, bartok2019regularized}

We also test the convergence of stress with respect to the radial cutoff of NAO basis functions.
In previous tests with the PBE functional,\cite{21CPC-Zheng} results using NAO basis converge to about 1 kBar at a radial cutoff of 8 a.u. and agree well with the results from the PW basis.
Here we perform the same test but with the SCAN functional and observe a similar pattern. As shown in Fig.~\ref{rcut_converg}, by choosing the radial cutoff to be 8 a.u., both DZP and TZDP basis sets yield converged results.

\subsection{Water Hexamer} 
Weak interactions, 
including the hydrogen bond (HB) and van der Waals interactions,
play a key role in the interactions between molecules.
It is well known that the use of GGA-level functionals within the framework of KSDFT largely overestimates the HB strength due to the self-interaction errors and inadequate treatment of the medium- and long-ranged van der Waals interactions,
which substantially undermines the accuracy of molecular interactions.\cite{14JCP-Distasio, 17NAS-Chen}
By incorporating the kinetic energy density in the XC functional,
the SCAN meta-GGA functional is able to capture the short- and intermediate-ranged van der Waals interactions and provides a more accurate description of hydrogen bonds.
Therefore, we first establish the accuracy of DFT calculations with the NAO basis set by evaluating the interaction strength between molecules and comparing the results with those obtained from the plane wave and Gaussian basis sets.
%
%
A typical example of  weakly interacting system is the water hexamer.\cite{16NC-Sun}
Here we choose four stable configurations of water hexamer, namely the prism, cage, book, and cyclic structures.\cite{08JCP-Santra, 98JCP-Kim, 02JCP-Losada}

\begin{table*}[htp]
\renewcommand{\arraystretch}{1.5}
\setlength{\tabcolsep}{2.4mm}
\caption{Interaction energies (in kcal/mol) for the dimers in S22 sets from the SCAN meta-GGA functional with the TZDP basis sets (using the ABACUS pacakge), the SCAN results (with the Gaussian basis and plane wave basis respectively) from Refs.~\onlinecite{15L-Sun, 16X-Peng} and the CCSD(T) results (with the Gaussian basis) from Ref.~\onlinecite{10PCCP-Podeszwa}.}
    \begin{center}
    \begin{tabular}{lccc}
    \hline
    \hline
          & SCAN (TZDP basis) & SCAN (Gaussian\cite{15L-Sun}, PW\cite{16X-Peng} basis) & CCSD(T) (Gaussian basis\cite{10PCCP-Podeszwa})\\
    \hline
    \multicolumn{4}{c}{7 hydrogen-bonded complexes}\\
    $\mathrm{NH_3}$ dimer ($\mathrm{C}_{2h}$) & 3.14 & 3.14, 3.12 & 3.15\\
    $\mathrm{H_2O}$ dimer ($\mathrm{C}_{s}$) & 5.44 & 5.39, 5.43 & 5.00\\
    Formic acid dimer ($\mathrm{C}_{2h}$) & 20.91 & 20.63, 20.93 & 18.75\\
    Formamide dimer ($\mathrm{C}_{2h}$) & 16.25 & 16.39, 16.54 & 16.06\\
    Uracil dimer ($\mathrm{C}_{2h}$) & 20.12 & 20.33, 20.49 & 20.64\\
    2-pyridone-2-aminopyridine ($\mathrm{C}_{1}$) & 16.76 & 16.69, 16.85 & 16.94\\
    Adenine-thymine WC ($\mathrm{C}_{1}$) & 15.68 & 15.88, 15.99 & 16.55\\
    \multicolumn{4}{c}{8 dispersion-bound complexes}\\
    $\mathrm{CH_4}$ dimer ($\mathrm{D}_{3d}$) & 0.38 & 0.37, 0.35 & 0.53\\
    $\mathrm{C_2H_4}$ dimer ($\mathrm{D}_{2d}$) & 1.03 & 1.07, 1.02 & 1.48\\
    Benzene-$\mathrm{CH_4}$ ($\mathrm{C}_{3}$) & 0.93 & 0.89, 0.87 & 1.45\\
    Benzene dimer ($\mathrm{C}_{2h}$) & 1.02 & 1.14, 1.07 & 2.66\\
    Pyrazine dimer ($\mathrm{C}_{s}$) & 2.74 & 2.71, 2.65 & 4.26\\
    Uracil dimer ($\mathrm{C}_{2}$) & 8.07 & 8.00, 7.96 & 9.78\\
    Indole-benzene ($\mathrm{C}_{1}$) & 2.09 & 2.19, 2.12 & 4.52\\
    Adenine-thymine ($\mathrm{C}_{1}$) & 8.83 & 8.69, 8.65 & 11.86\\
    \multicolumn{4}{c}{7 mixed complexes}\\
    $\mathrm{C_2H_4}$-$\mathrm{C_2H_2}$ ($\mathrm{C}_{2v}$) & 1.40 & 1.35, 1.34 & 1.50\\
    Benzene-$\mathrm{H_2O}$ ($\mathrm{C}_{s}$) & 3.41 & 3.30, 3.28 & 3.28\\
    Benzene-$\mathrm{NH_3}$ ($\mathrm{C}_{s}$) & 2.07 & 2.00, 1.99 & 2.32\\
    Benzene-HCN ($\mathrm{C}_{s}$) & 4.28 & 4.08, 4.06 & 4.54\\
    Benzene dimer ($\mathrm{C}_{2v}$) & 1.58 & 1.50, 1.48 & 2.72\\
    Indole-benzene ($\mathrm{C}_{s}$) & 4.26 & 4.07, 4.07 & 5.63\\
    Phenol dimer ($\mathrm{C}_{1}$) & 5.96 & 5.91, 5.91 & 7.10\\
        \hline
       
    \end{tabular}
    \end{center}
    \label{tab:S22}
\end{table*}

\begin{table*}[htp]
\renewcommand{\arraystretch}{1.5}
\setlength{\tabcolsep}{1.5mm}
\caption{Lattice constants ($a$ in \AA) and bulk moduli ($B$ in GPa) calculated with NAO (DZP and TZDP) and PW basis sets.
The experiment results (Exp) extrapolated to 0 K are cited from Ref.~\onlinecite{10B-Harl} and the reference data (Ref) are cited from Ref.~\onlinecite{16X-Peng}.}
    \begin{center}
    \begin{tabular}{lccccccccc}
    \hline
    \hline
          & \multicolumn{2}{c}{DZP} & \multicolumn{2}{c}{TZDP} & \multicolumn{2}{c}{PW} & \multicolumn{2}{c}{Exp} & Ref  \\
        & $a$ & $B$ & $a$ & $B$ & $a$ & $B$ & $a$ & $B$  & $a$\\
        \hline
       Diamond & 3.56 & 444 & 3.56 & 454 & 3.56 & 449 & 3.55 & 443 & 3.55 \\
       Si   & 5.45 & 92  & 5.45 & 91  & 5.45 & 94  & 5.42 & 99  & 5.43 \\
       Ge   & 5.58 & 73  & 5.58 & 73  & 5.59 & 77  & 5.64 & 76  & 5.66 \\
       SiC  & 4.38 & 218 & 4.37 & 218 & 4.36 & 220 & 4.35 & 225 & 4.35 \\
       AlN  & 4.39 & 218 & 4.37 & 215 & 4.36 & 219 & 4.37 & 202 & 4.36 \\
       AlP  & 5.48 & 91  & 5.48 & 93  & 5.47 & 93  & 5.45 & 86  & 5.47 \\
       AlAs & 5.69 & 78  & 5.69 & 78  & 5.68 & 78  & 5.65 & 77  & 5.67 \\
       GaN  & 4.49 & 179 & 4.49 & 181 & 4.48 & 181 & 4.52 & 210 & 4.50 \\
       GaP  & 5.42 & 90  & 5.42 & 91  & 5.41 & 81  & 5.44 & 89  & 5.45 \\
       GaAs & 5.66 & 69  & 5.65 & 69  & 5.66 & 67  & 5.64 & 76  & 5.66 \\
       InP  & 5.88 & 68  & 5.87 & 68  & 5.86 & 69  & 5.86 & 71  & 5.89 \\
       InAs & 6.11 & 55  & 6.10 & 56  & 6.10 & 56  & 6.05 & 58  & 6.09 \\
       InSb & 6.46 & 44  & 6.45 & 44  & 6.45 & 44  & 6.47 & 46  & 6.52 \\
    \hline
    \end{tabular}
    \end{center}
    \label{tab:semicon}
\end{table*}

High-precision quantum chemistry methods such as the coupled cluster singles and doubles with perturbative triples (CCSD(T)) and M$\mathrm{\Ddot{o}}$ller-Plesset perturbation theory (MP2)\cite{34PR-Moller} predict the cyclic structure of water hexamer to be the most stable structure with the lowest energy, followed by the book, cage, and prism structures.\cite{08JCP-Santra}
However, most GGA-level functionals and hybrid functionals fail to predict a correct energy ordering for the four water hexamers due to the lack of a proper description of the van der Waals interactions.\cite{08JCP-Santra}
On the contrary, the SCAN functional has been shown to correctly predict the order of water hexamers with the Gaussian basis extrapolated to the complete basis set limit.\cite{16NC-Sun} 
By utilizing the ABACUS package with the newly implemented SCAN function, we calculate the energies of the four water hexamer clusters with the TZDP NAO basis set, 
and align the results by setting the total energy of the prism structure to 0, and the results are shown in Fig.~\ref{water_cluster}. 
We find the SCAN functional with NAO basis sets predicts the correct energy ordering and agrees well with the results listed in Ref.~\onlinecite{16NC-Sun}.
We also perform basis set extrapolation, and the difference of calculated energies between the complete basis limit and the TZDP basis is less than 3 meV per water molecule. Therefore, we conclude that our results are reliable.

\subsection{Weakly Interacting Molecules}

We also test the accuracy of the SCAN functional on a wider range of weakly interacting systems in the S22 dataset\cite{06PCCP-Petr}. Here we adopt the NAO basis sets provided by ABACUS.
The dataset contains three different groups of weakly interacting molecular systems, where the interactions are dominated by hydrogen bonds, dispersion interactions, and mixed interactions, respectively.
We calculate the interaction energies with the SCAN functional and the TZDP basis set by using the ABACUS package; the results are 
listed in Table~\ref{tab:S22}.
Basis set superposition error (BSSE) is corrected using the counterpoise method in the computation.\cite{70MP-Boys}
We also list the SCAN results with the Gaussian basis\cite{15L-Sun}. In addition, we include the SCAN results\cite{16X-Peng} calculated from the plane wave basis within the framework of the projector augmented wave (PAW) method\cite{94B-Blochl}, as well as the CCSD(T) results with the Gaussian basis set\cite{10PCCP-Podeszwa}.
The resulting interaction energy errors with respect to these reference SCAN data are within 0.3 kcal/mol, most of which are within 0.1 kcal/mol, proving that the accuracy of the SCAN functional with the TZDP basis set is sufficient for the description of the weak interactions between molecules.
%
The largest deviation (about 0.2 kcal/mol) comes from the case of benzene-HCN dimer and indole-benzene dimer.
However, we note that the deviation between the SCAN functional with the PW basis and the CCSD(T) method with the Gaussian basis could be as large as 0.3 kcal/mol (in the case of formic acid dimer), indicating this level of deviation is expected when comparing results using different basis sets.

\subsection{Semiconductors}
Besides the molecular systems, we also take 13 semiconductors as examples to test the SCAN functional with both NAO and PW basis.
For the semiconductors,
we compute the lattice constants $a$ and bulk moduli $B$ with the DZP and TZDP NAO basis, as well as the plane wave basis. The data are listed in Table~\ref{tab:semicon} and compared with Ref.~\onlinecite{16X-Peng} and experiment\cite{10B-Harl}.
%
%
The bulk moduli $B$ are calculated by 
\begin{equation}
    B=V\frac{\partial^2E}{\partial V^2}|_{V=V_0},
\end{equation}
where $V$ is the volume per atom, $E$ depicts the energy per atom, and $V_0$ denotes the volume per atom that minimize the energy.
%
%
By comparing the NAO results with those from the plane wave basis, it could be seen that the calculated lattice constants have converged to within 0.01 \AA~for most semiconductors at the DZP basis set level, except for a few cases, such as SiC, AlN, and InP.
For bulk moduli, most semiconductors converge to within 1 GPa at the DZP level,
except for a few cases like diamond and AlP.
Compared with the referenced data calculated using the PAW method,\cite{16X-Peng}
the lattice constants from the NAO basis mostly fall within 0.03 \AA~around the reference results, except for Ge and InSb, where the errors are 0.06 and 0.08 \AA\, respectively.
The calculated bulk moduli also show satisfactory agreement with the experiment,
most of which fall into 10 GPa around the experiment results, proving that SCAN functional produces highly accurate results in the simulations of bulk semiconductor materials.
Therefore, we conclude that results with NAO basis are consistent with the PW basis in the simulation of bulk systems using the SCAN meta-GGA functional.
We also show that the accuracy of NAO simulations at the DZP level is sufficient for most of the tested systems. In general, TZDP is needed only for cases where high accuracy is required.

\section{Conclusions}\label{sec4}
In conclusion, we implemented meta-GGA functionals in both  numerical atomic orbital and plane-wave basis sets in the electronic structure software ABACUS. 
We adopted the formulation commonly referred to as ODDM\cite{13JCP-Zahariev} or FDO\cite{16JCP-Womack} method rather than the numerically challenging OEP method. 
Apart from the $\tau$-dependent contribution to the Hamiltonian matrix, we also implemented the force and stress with both basis sets. 

To validate our stress implementation, we compared the stress calculated using the finite difference method with the analytic results and obtained a satisfactory agreement between the two methods. The convergence of the results with respect to the basis sets was also discussed. We then calculated the binding energies of water hexamers, interaction energies of weak interacting molecules in the S22 dataset, as well as lattice constants and bulk moduli of 13 semiconductors. The results were consistent with the experimental and previous computational values.
We expect our implementation of the meta-GGA exchange-correlation functionals in ABACUS will facilitate more method developments and applications in the future. 


\section*{Acknowledgement}
We thank Tianqi Zhao, Xingliang Peng, Chun Cai, Qi Ou, and Xiaokuang Bai for improving the ABACUS package from various aspects, which include adding test examples and setting up environments and workflows. We also thank Jianwei Sun for discussions on the meta-GGA functional and helpful suggestions on the manuscript.
The work of M.C., R.L., and X.L. was supported by the National Science Foundation of China under Grant
No. 12122401, 12074007, and 12135002. The authors gratefully acknowledge funding support from the AI for Science Institute, Beijing (AISI). Computational work in this study benefits from the use of the high-performance computing platform of Peking University and the Bohrium platform supported by DP Technology.

\bibliography{main}

\begin{thebibliography}{69}
\expandafter\ifx\csname natexlab\endcsname\relax\def\natexlab#1{#1}\fi
\expandafter\ifx\csname bibnamefont\endcsname\relax
  \def\bibnamefont#1{#1}\fi
\expandafter\ifx\csname bibfnamefont\endcsname\relax
  \def\bibfnamefont#1{#1}\fi
\expandafter\ifx\csname citenamefont\endcsname\relax
  \def\citenamefont#1{#1}\fi
\expandafter\ifx\csname url\endcsname\relax
  \def\url#1{\texttt{#1}}\fi
\expandafter\ifx\csname urlprefix\endcsname\relax\def\urlprefix{URL }\fi
\providecommand{\bibinfo}[2]{#2}
\providecommand{\eprint}[2][]{\url{#2}}

\bibitem[{\citenamefont{Hohenberg and Kohn}(1964)}]{64PR-Hohenberg}
\bibinfo{author}{\bibfnamefont{P.}~\bibnamefont{Hohenberg}} \bibnamefont{and}
  \bibinfo{author}{\bibfnamefont{W.}~\bibnamefont{Kohn}},
  \bibinfo{journal}{Phys. \ Rev.} \textbf{\bibinfo{volume}{136}},
  \bibinfo{pages}{864B} (\bibinfo{year}{1964}).

\bibitem[{\citenamefont{Kohn and Sham}(1965)}]{65PR-Kohn}
\bibinfo{author}{\bibfnamefont{W.}~\bibnamefont{Kohn}} \bibnamefont{and}
  \bibinfo{author}{\bibfnamefont{L.~J.} \bibnamefont{Sham}},
  \bibinfo{journal}{Phys. \ Rev.} \textbf{\bibinfo{volume}{140}},
  \bibinfo{pages}{1133A} (\bibinfo{year}{1965}).

\bibitem[{\citenamefont{Albert et~al.}(2011)\citenamefont{Albert, Ivanov,
  Tretiak, and Kilina}}]{11JCPC-Albert}
\bibinfo{author}{\bibfnamefont{V.~V.} \bibnamefont{Albert}},
  \bibinfo{author}{\bibfnamefont{S.~A.} \bibnamefont{Ivanov}},
  \bibinfo{author}{\bibfnamefont{S.}~\bibnamefont{Tretiak}}, \bibnamefont{and}
  \bibinfo{author}{\bibfnamefont{S.~V.} \bibnamefont{Kilina}},
  \bibinfo{journal}{The Journal of Physical Chemistry C}
  \textbf{\bibinfo{volume}{115}}, \bibinfo{pages}{15793}
  (\bibinfo{year}{2011}).

\bibitem[{\citenamefont{Perdew et~al.}(2005)\citenamefont{Perdew, Ruzsinszky,
  Tao, Staroverov, Scuseria, and Csonka}}]{05JCP-Perdew}
\bibinfo{author}{\bibfnamefont{J.~P.} \bibnamefont{Perdew}},
  \bibinfo{author}{\bibfnamefont{A.}~\bibnamefont{Ruzsinszky}},
  \bibinfo{author}{\bibfnamefont{J.}~\bibnamefont{Tao}},
  \bibinfo{author}{\bibfnamefont{V.~N.} \bibnamefont{Staroverov}},
  \bibinfo{author}{\bibfnamefont{G.~E.} \bibnamefont{Scuseria}},
  \bibnamefont{and} \bibinfo{author}{\bibfnamefont{G.~I.}
  \bibnamefont{Csonka}}, \bibinfo{journal}{The Journal of chemical physics}
  \textbf{\bibinfo{volume}{123}}, \bibinfo{pages}{062201}
  (\bibinfo{year}{2005}).

\bibitem[{\citenamefont{Womack et~al.}(2016)\citenamefont{Womack, Mardirossian,
  Head-Gordon, and Skylaris}}]{16JCP-Womack}
\bibinfo{author}{\bibfnamefont{J.~C.} \bibnamefont{Womack}},
  \bibinfo{author}{\bibfnamefont{N.}~\bibnamefont{Mardirossian}},
  \bibinfo{author}{\bibfnamefont{M.}~\bibnamefont{Head-Gordon}},
  \bibnamefont{and} \bibinfo{author}{\bibfnamefont{C.-K.}
  \bibnamefont{Skylaris}}, \bibinfo{journal}{The Journal of Chemical Physics}
  \textbf{\bibinfo{volume}{145}}, \bibinfo{pages}{204114}
  (\bibinfo{year}{2016}).

\bibitem[{\citenamefont{Hao et~al.}(2013)\citenamefont{Hao, Sun, Xiao,
  Ruzsinszky, Csonka, Tao, Glindmeyer, and Perdew}}]{13JCTC-Hao}
\bibinfo{author}{\bibfnamefont{P.}~\bibnamefont{Hao}},
  \bibinfo{author}{\bibfnamefont{J.}~\bibnamefont{Sun}},
  \bibinfo{author}{\bibfnamefont{B.}~\bibnamefont{Xiao}},
  \bibinfo{author}{\bibfnamefont{A.}~\bibnamefont{Ruzsinszky}},
  \bibinfo{author}{\bibfnamefont{G.~I.} \bibnamefont{Csonka}},
  \bibinfo{author}{\bibfnamefont{J.}~\bibnamefont{Tao}},
  \bibinfo{author}{\bibfnamefont{S.}~\bibnamefont{Glindmeyer}},
  \bibnamefont{and} \bibinfo{author}{\bibfnamefont{J.~P.}
  \bibnamefont{Perdew}}, \bibinfo{journal}{Journal of Chemical Theory and
  Computation} \textbf{\bibinfo{volume}{9}}, \bibinfo{pages}{355}
  (\bibinfo{year}{2013}).

\bibitem[{\citenamefont{Tao et~al.}(2003{\natexlab{a}})\citenamefont{Tao,
  Perdew, Staroverov, and Scuseria}}]{2003APS-Tao}
\bibinfo{author}{\bibfnamefont{J.}~\bibnamefont{Tao}},
  \bibinfo{author}{\bibfnamefont{J.~P.} \bibnamefont{Perdew}},
  \bibinfo{author}{\bibfnamefont{V.~N.} \bibnamefont{Staroverov}},
  \bibnamefont{and} \bibinfo{author}{\bibfnamefont{G.~E.}
  \bibnamefont{Scuseria}}, \bibinfo{journal}{Physical Review Letters}
  \textbf{\bibinfo{volume}{91}}, \bibinfo{pages}{146401}
  (\bibinfo{year}{2003}{\natexlab{a}}).

\bibitem[{\citenamefont{Sun et~al.}(2011{\natexlab{a}})\citenamefont{Sun,
  Marsman, Ruzsinszky, Kresse, and Perdew}}]{2011APS-Sun}
\bibinfo{author}{\bibfnamefont{J.}~\bibnamefont{Sun}},
  \bibinfo{author}{\bibfnamefont{M.}~\bibnamefont{Marsman}},
  \bibinfo{author}{\bibfnamefont{A.}~\bibnamefont{Ruzsinszky}},
  \bibinfo{author}{\bibfnamefont{G.}~\bibnamefont{Kresse}}, \bibnamefont{and}
  \bibinfo{author}{\bibfnamefont{J.~P.} \bibnamefont{Perdew}},
  \bibinfo{journal}{Physical Review B} \textbf{\bibinfo{volume}{83}},
  \bibinfo{pages}{121410} (\bibinfo{year}{2011}{\natexlab{a}}).

\bibitem[{\citenamefont{Sun et~al.}(2013)\citenamefont{Sun, Xiao, Fang,
  Haunschild, Hao, Ruzsinszky, Csonka, Scuseria, and Perdew}}]{13L-Sun}
\bibinfo{author}{\bibfnamefont{J.}~\bibnamefont{Sun}},
  \bibinfo{author}{\bibfnamefont{B.}~\bibnamefont{Xiao}},
  \bibinfo{author}{\bibfnamefont{Y.}~\bibnamefont{Fang}},
  \bibinfo{author}{\bibfnamefont{R.}~\bibnamefont{Haunschild}},
  \bibinfo{author}{\bibfnamefont{P.}~\bibnamefont{Hao}},
  \bibinfo{author}{\bibfnamefont{A.}~\bibnamefont{Ruzsinszky}},
  \bibinfo{author}{\bibfnamefont{G.~I.} \bibnamefont{Csonka}},
  \bibinfo{author}{\bibfnamefont{G.~E.} \bibnamefont{Scuseria}},
  \bibnamefont{and} \bibinfo{author}{\bibfnamefont{J.~P.}
  \bibnamefont{Perdew}}, \bibinfo{journal}{Physical Review Letters}
  \textbf{\bibinfo{volume}{111}}, \bibinfo{pages}{106401}
  (\bibinfo{year}{2013}).

\bibitem[{\citenamefont{Sun et~al.}(2015)\citenamefont{Sun, Ruzsinszky, and
  Perdew}}]{15L-Sun}
\bibinfo{author}{\bibfnamefont{J.}~\bibnamefont{Sun}},
  \bibinfo{author}{\bibfnamefont{A.}~\bibnamefont{Ruzsinszky}},
  \bibnamefont{and} \bibinfo{author}{\bibfnamefont{J.~P.}
  \bibnamefont{Perdew}}, \bibinfo{journal}{Physical Review Letters}
  \textbf{\bibinfo{volume}{115}}, \bibinfo{pages}{036402}
  (\bibinfo{year}{2015}).

\bibitem[{\citenamefont{Zhang et~al.}(2018{\natexlab{a}})\citenamefont{Zhang,
  Kitchaev, Yang, Chen, Dacek, Sarmiento-P{\'e}rez, Marques, Peng, Ceder,
  Perdew et~al.}}]{18NPJ-Zhang}
\bibinfo{author}{\bibfnamefont{Y.}~\bibnamefont{Zhang}},
  \bibinfo{author}{\bibfnamefont{D.~A.} \bibnamefont{Kitchaev}},
  \bibinfo{author}{\bibfnamefont{J.}~\bibnamefont{Yang}},
  \bibinfo{author}{\bibfnamefont{T.}~\bibnamefont{Chen}},
  \bibinfo{author}{\bibfnamefont{S.~T.} \bibnamefont{Dacek}},
  \bibinfo{author}{\bibfnamefont{R.~A.} \bibnamefont{Sarmiento-P{\'e}rez}},
  \bibinfo{author}{\bibfnamefont{M.~A.} \bibnamefont{Marques}},
  \bibinfo{author}{\bibfnamefont{H.}~\bibnamefont{Peng}},
  \bibinfo{author}{\bibfnamefont{G.}~\bibnamefont{Ceder}},
  \bibinfo{author}{\bibfnamefont{J.~P.} \bibnamefont{Perdew}},
  \bibnamefont{et~al.}, \bibinfo{journal}{npj Computational Materials}
  \textbf{\bibinfo{volume}{4}}, \bibinfo{pages}{1}
  (\bibinfo{year}{2018}{\natexlab{a}}).

\bibitem[{\citenamefont{Zhang et~al.}(2018{\natexlab{b}})\citenamefont{Zhang,
  Reilly, Tkatchenko, and Scheffler}}]{18NJP-Zhang}
\bibinfo{author}{\bibfnamefont{G.-X.} \bibnamefont{Zhang}},
  \bibinfo{author}{\bibfnamefont{A.~M.} \bibnamefont{Reilly}},
  \bibinfo{author}{\bibfnamefont{A.}~\bibnamefont{Tkatchenko}},
  \bibnamefont{and}
  \bibinfo{author}{\bibfnamefont{M.}~\bibnamefont{Scheffler}},
  \bibinfo{journal}{New Journal of Physics} \textbf{\bibinfo{volume}{20}},
  \bibinfo{pages}{063020} (\bibinfo{year}{2018}{\natexlab{b}}).

\bibitem[{\citenamefont{Kingsbury et~al.}(2022)\citenamefont{Kingsbury, Gupta,
  Bartel, Munro, Dwaraknath, Horton, and Persson}}]{22PRM-Kingsbury}
\bibinfo{author}{\bibfnamefont{R.}~\bibnamefont{Kingsbury}},
  \bibinfo{author}{\bibfnamefont{A.~S.} \bibnamefont{Gupta}},
  \bibinfo{author}{\bibfnamefont{C.~J.} \bibnamefont{Bartel}},
  \bibinfo{author}{\bibfnamefont{J.~M.} \bibnamefont{Munro}},
  \bibinfo{author}{\bibfnamefont{S.}~\bibnamefont{Dwaraknath}},
  \bibinfo{author}{\bibfnamefont{M.}~\bibnamefont{Horton}}, \bibnamefont{and}
  \bibinfo{author}{\bibfnamefont{K.~A.} \bibnamefont{Persson}},
  \bibinfo{journal}{Physical Review Materials} \textbf{\bibinfo{volume}{6}},
  \bibinfo{pages}{013801} (\bibinfo{year}{2022}).

\bibitem[{\citenamefont{Shahi et~al.}(2018)\citenamefont{Shahi, Sun, and
  Perdew}}]{18PRB-Shahi}
\bibinfo{author}{\bibfnamefont{C.}~\bibnamefont{Shahi}},
  \bibinfo{author}{\bibfnamefont{J.}~\bibnamefont{Sun}}, \bibnamefont{and}
  \bibinfo{author}{\bibfnamefont{J.~P.} \bibnamefont{Perdew}},
  \bibinfo{journal}{Physical Review B} \textbf{\bibinfo{volume}{97}},
  \bibinfo{pages}{094111} (\bibinfo{year}{2018}).

\bibitem[{\citenamefont{Sun et~al.}(2016)\citenamefont{Sun, Remsing, Zhang,
  Sun, Ruzsinszky, Peng, Yang, Paul, Waghmare, Wu et~al.}}]{16NC-Sun}
\bibinfo{author}{\bibfnamefont{J.}~\bibnamefont{Sun}},
  \bibinfo{author}{\bibfnamefont{R.~C.} \bibnamefont{Remsing}},
  \bibinfo{author}{\bibfnamefont{Y.}~\bibnamefont{Zhang}},
  \bibinfo{author}{\bibfnamefont{Z.}~\bibnamefont{Sun}},
  \bibinfo{author}{\bibfnamefont{A.}~\bibnamefont{Ruzsinszky}},
  \bibinfo{author}{\bibfnamefont{H.}~\bibnamefont{Peng}},
  \bibinfo{author}{\bibfnamefont{Z.}~\bibnamefont{Yang}},
  \bibinfo{author}{\bibfnamefont{A.}~\bibnamefont{Paul}},
  \bibinfo{author}{\bibfnamefont{U.}~\bibnamefont{Waghmare}},
  \bibinfo{author}{\bibfnamefont{X.}~\bibnamefont{Wu}}, \bibnamefont{et~al.},
  \bibinfo{journal}{Nature Chemistry} \textbf{\bibinfo{volume}{8}},
  \bibinfo{pages}{831} (\bibinfo{year}{2016}).

\bibitem[{\citenamefont{Chen et~al.}(2017)\citenamefont{Chen, Ko, Remsing,
  Calegari~Andrade, Santra, Sun, Selloni, Car, Klein, Perdew
  et~al.}}]{17NAS-Chen}
\bibinfo{author}{\bibfnamefont{M.}~\bibnamefont{Chen}},
  \bibinfo{author}{\bibfnamefont{H.-Y.} \bibnamefont{Ko}},
  \bibinfo{author}{\bibfnamefont{R.~C.} \bibnamefont{Remsing}},
  \bibinfo{author}{\bibfnamefont{M.~F.} \bibnamefont{Calegari~Andrade}},
  \bibinfo{author}{\bibfnamefont{B.}~\bibnamefont{Santra}},
  \bibinfo{author}{\bibfnamefont{Z.}~\bibnamefont{Sun}},
  \bibinfo{author}{\bibfnamefont{A.}~\bibnamefont{Selloni}},
  \bibinfo{author}{\bibfnamefont{R.}~\bibnamefont{Car}},
  \bibinfo{author}{\bibfnamefont{M.~L.} \bibnamefont{Klein}},
  \bibinfo{author}{\bibfnamefont{J.~P.} \bibnamefont{Perdew}},
  \bibnamefont{et~al.}, \bibinfo{journal}{Proceedings of the National Academy
  of Sciences} \textbf{\bibinfo{volume}{114}}, \bibinfo{pages}{10846}
  (\bibinfo{year}{2017}).

\bibitem[{\citenamefont{LaCount and Gygi}(2019)}]{19JCP-Lacount}
\bibinfo{author}{\bibfnamefont{M.~D.} \bibnamefont{LaCount}} \bibnamefont{and}
  \bibinfo{author}{\bibfnamefont{F.}~\bibnamefont{Gygi}}, \bibinfo{journal}{The
  Journal of Chemical Physics} \textbf{\bibinfo{volume}{151}},
  \bibinfo{pages}{164101} (\bibinfo{year}{2019}).

\bibitem[{\citenamefont{Liu et~al.}(2022)\citenamefont{Liu, Zhang, Liang, Liu,
  Wu, and Chen}}]{22JCP-Liu}
\bibinfo{author}{\bibfnamefont{R.}~\bibnamefont{Liu}},
  \bibinfo{author}{\bibfnamefont{C.}~\bibnamefont{Zhang}},
  \bibinfo{author}{\bibfnamefont{X.}~\bibnamefont{Liang}},
  \bibinfo{author}{\bibfnamefont{J.}~\bibnamefont{Liu}},
  \bibinfo{author}{\bibfnamefont{X.}~\bibnamefont{Wu}}, \bibnamefont{and}
  \bibinfo{author}{\bibfnamefont{M.}~\bibnamefont{Chen}}, \bibinfo{journal}{The
  Journal of Chemical Physics} \textbf{\bibinfo{volume}{157}},
  \bibinfo{pages}{024503} (\bibinfo{year}{2022}).

\bibitem[{\citenamefont{Liu et~al.}(2023)\citenamefont{Liu, Liu, Cao, and
  Chen}}]{23PCCP-Liu}
\bibinfo{author}{\bibfnamefont{J.}~\bibnamefont{Liu}},
  \bibinfo{author}{\bibfnamefont{R.}~\bibnamefont{Liu}},
  \bibinfo{author}{\bibfnamefont{Y.}~\bibnamefont{Cao}}, \bibnamefont{and}
  \bibinfo{author}{\bibfnamefont{M.}~\bibnamefont{Chen}},
  \bibinfo{journal}{Physical Chemistry Chemical Physics}
  \textbf{\bibinfo{volume}{25}}, \bibinfo{pages}{983} (\bibinfo{year}{2023}).

\bibitem[{\citenamefont{Zhang et~al.}(2022)\citenamefont{Zhang, Yue,
  Panagiotopoulos, Klein, and Wu}}]{22NC-Zhang}
\bibinfo{author}{\bibfnamefont{C.}~\bibnamefont{Zhang}},
  \bibinfo{author}{\bibfnamefont{S.}~\bibnamefont{Yue}},
  \bibinfo{author}{\bibfnamefont{A.~Z.} \bibnamefont{Panagiotopoulos}},
  \bibinfo{author}{\bibfnamefont{M.~L.} \bibnamefont{Klein}}, \bibnamefont{and}
  \bibinfo{author}{\bibfnamefont{X.}~\bibnamefont{Wu}},
  \bibinfo{journal}{Nature Communications} \textbf{\bibinfo{volume}{13}},
  \bibinfo{pages}{822} (\bibinfo{year}{2022}).

\bibitem[{\citenamefont{Yang et~al.}(2019)\citenamefont{Yang, Kitchaev, and
  Ceder}}]{19PRB-Yang}
\bibinfo{author}{\bibfnamefont{J.~H.} \bibnamefont{Yang}},
  \bibinfo{author}{\bibfnamefont{D.~A.} \bibnamefont{Kitchaev}},
  \bibnamefont{and} \bibinfo{author}{\bibfnamefont{G.}~\bibnamefont{Ceder}},
  \bibinfo{journal}{Physical Review B} \textbf{\bibinfo{volume}{100}},
  \bibinfo{pages}{035132} (\bibinfo{year}{2019}).

\bibitem[{\citenamefont{Kanungo et~al.}(2021)\citenamefont{Kanungo, Zimmerman,
  and Gavini}}]{kanungo2021comparison}
\bibinfo{author}{\bibfnamefont{B.}~\bibnamefont{Kanungo}},
  \bibinfo{author}{\bibfnamefont{P.~M.} \bibnamefont{Zimmerman}},
  \bibnamefont{and} \bibinfo{author}{\bibfnamefont{V.}~\bibnamefont{Gavini}},
  \bibinfo{journal}{The Journal of Physical Chemistry Letters}
  \textbf{\bibinfo{volume}{12}}, \bibinfo{pages}{12012} (\bibinfo{year}{2021}).

\bibitem[{\citenamefont{Zhang et~al.}(2021)\citenamefont{Zhang, Tang, Chen, Xu,
  Zhang, Qiu, Perdew, Klein, and Wu}}]{zhang2021modeling}
\bibinfo{author}{\bibfnamefont{C.}~\bibnamefont{Zhang}},
  \bibinfo{author}{\bibfnamefont{F.}~\bibnamefont{Tang}},
  \bibinfo{author}{\bibfnamefont{M.}~\bibnamefont{Chen}},
  \bibinfo{author}{\bibfnamefont{J.}~\bibnamefont{Xu}},
  \bibinfo{author}{\bibfnamefont{L.}~\bibnamefont{Zhang}},
  \bibinfo{author}{\bibfnamefont{D.~Y.} \bibnamefont{Qiu}},
  \bibinfo{author}{\bibfnamefont{J.~P.} \bibnamefont{Perdew}},
  \bibinfo{author}{\bibfnamefont{M.~L.} \bibnamefont{Klein}}, \bibnamefont{and}
  \bibinfo{author}{\bibfnamefont{X.}~\bibnamefont{Wu}}, \bibinfo{journal}{The
  Journal of Physical Chemistry B} \textbf{\bibinfo{volume}{125}},
  \bibinfo{pages}{11444} (\bibinfo{year}{2021}).

\bibitem[{\citenamefont{Sun et~al.}(2011{\natexlab{b}})\citenamefont{Sun,
  Marsman, Csonka, Ruzsinszky, Hao, Kim, Kresse, and Perdew}}]{11B-Sun}
\bibinfo{author}{\bibfnamefont{J.}~\bibnamefont{Sun}},
  \bibinfo{author}{\bibfnamefont{M.}~\bibnamefont{Marsman}},
  \bibinfo{author}{\bibfnamefont{G.~I.} \bibnamefont{Csonka}},
  \bibinfo{author}{\bibfnamefont{A.}~\bibnamefont{Ruzsinszky}},
  \bibinfo{author}{\bibfnamefont{P.}~\bibnamefont{Hao}},
  \bibinfo{author}{\bibfnamefont{Y.-S.} \bibnamefont{Kim}},
  \bibinfo{author}{\bibfnamefont{G.}~\bibnamefont{Kresse}}, \bibnamefont{and}
  \bibinfo{author}{\bibfnamefont{J.~P.} \bibnamefont{Perdew}},
  \bibinfo{journal}{Physical Review B} \textbf{\bibinfo{volume}{84}},
  \bibinfo{pages}{035117} (\bibinfo{year}{2011}{\natexlab{b}}).

\bibitem[{\citenamefont{Yao and Kanai}(2017)}]{17JCP-Yao}
\bibinfo{author}{\bibfnamefont{Y.}~\bibnamefont{Yao}} \bibnamefont{and}
  \bibinfo{author}{\bibfnamefont{Y.}~\bibnamefont{Kanai}},
  \bibinfo{journal}{The Journal of Chemical Physics}
  \textbf{\bibinfo{volume}{146}}, \bibinfo{pages}{224105}
  (\bibinfo{year}{2017}).

\bibitem[{\citenamefont{Neumann et~al.}(1996)\citenamefont{Neumann, Nobes, and
  Handy}}]{96MP-Neumann}
\bibinfo{author}{\bibfnamefont{R.}~\bibnamefont{Neumann}},
  \bibinfo{author}{\bibfnamefont{R.~H.} \bibnamefont{Nobes}}, \bibnamefont{and}
  \bibinfo{author}{\bibfnamefont{N.~C.} \bibnamefont{Handy}},
  \bibinfo{journal}{Molecular Physics} \textbf{\bibinfo{volume}{87}},
  \bibinfo{pages}{1} (\bibinfo{year}{1996}).

\bibitem[{\citenamefont{Tao et~al.}(2003{\natexlab{b}})\citenamefont{Tao,
  Perdew, Staroverov, and Scuseria}}]{03L-Tao}
\bibinfo{author}{\bibfnamefont{J.}~\bibnamefont{Tao}},
  \bibinfo{author}{\bibfnamefont{J.~P.} \bibnamefont{Perdew}},
  \bibinfo{author}{\bibfnamefont{V.~N.} \bibnamefont{Staroverov}},
  \bibnamefont{and} \bibinfo{author}{\bibfnamefont{G.~E.}
  \bibnamefont{Scuseria}}, \bibinfo{journal}{Physical Review Letters}
  \textbf{\bibinfo{volume}{91}}, \bibinfo{pages}{146401}
  (\bibinfo{year}{2003}{\natexlab{b}}).

\bibitem[{\citenamefont{Chen et~al.}(2010)\citenamefont{Chen, Guo, and
  He}}]{10JPCM-Chen}
\bibinfo{author}{\bibfnamefont{M.}~\bibnamefont{Chen}},
  \bibinfo{author}{\bibfnamefont{G.-C.} \bibnamefont{Guo}}, \bibnamefont{and}
  \bibinfo{author}{\bibfnamefont{L.}~\bibnamefont{He}},
  \bibinfo{journal}{Journal of Physics: Condensed Matter}
  \textbf{\bibinfo{volume}{22}}, \bibinfo{pages}{445501}
  (\bibinfo{year}{2010}).

\bibitem[{\citenamefont{Chen et~al.}(2011)\citenamefont{Chen, Guo, and
  He}}]{11JPCM-Chen}
\bibinfo{author}{\bibfnamefont{M.}~\bibnamefont{Chen}},
  \bibinfo{author}{\bibfnamefont{G.-C.} \bibnamefont{Guo}}, \bibnamefont{and}
  \bibinfo{author}{\bibfnamefont{L.}~\bibnamefont{He}},
  \bibinfo{journal}{Journal of Physics: Condensed Matter}
  \textbf{\bibinfo{volume}{23}}, \bibinfo{pages}{325501}
  (\bibinfo{year}{2011}).

\bibitem[{\citenamefont{Anglada et~al.}(2002)\citenamefont{Anglada, Soler,
  Junquera, and Artacho}}]{01PRB-Anglada}
\bibinfo{author}{\bibfnamefont{E.}~\bibnamefont{Anglada}},
  \bibinfo{author}{\bibfnamefont{J.~M.} \bibnamefont{Soler}},
  \bibinfo{author}{\bibfnamefont{J.}~\bibnamefont{Junquera}}, \bibnamefont{and}
  \bibinfo{author}{\bibfnamefont{E.}~\bibnamefont{Artacho}},
  \bibinfo{journal}{Physical Review B} \textbf{\bibinfo{volume}{66}},
  \bibinfo{pages}{205101} (\bibinfo{year}{2002}).

\bibitem[{\citenamefont{Junquera et~al.}(2001)\citenamefont{Junquera, Paz,
  S{\'a}nchez-Portal, and Artacho}}]{01PRB-Junquera}
\bibinfo{author}{\bibfnamefont{J.}~\bibnamefont{Junquera}},
  \bibinfo{author}{\bibfnamefont{{\'O}.}~\bibnamefont{Paz}},
  \bibinfo{author}{\bibfnamefont{D.}~\bibnamefont{S{\'a}nchez-Portal}},
  \bibnamefont{and} \bibinfo{author}{\bibfnamefont{E.}~\bibnamefont{Artacho}},
  \bibinfo{journal}{Physical Review B} \textbf{\bibinfo{volume}{64}},
  \bibinfo{pages}{235111} (\bibinfo{year}{2001}).

\bibitem[{\citenamefont{Shang et~al.}(2010)\citenamefont{Shang, Xiang, Li, and
  Yang}}]{10IRPC-Shang}
\bibinfo{author}{\bibfnamefont{H.}~\bibnamefont{Shang}},
  \bibinfo{author}{\bibfnamefont{H.}~\bibnamefont{Xiang}},
  \bibinfo{author}{\bibfnamefont{Z.}~\bibnamefont{Li}}, \bibnamefont{and}
  \bibinfo{author}{\bibfnamefont{J.}~\bibnamefont{Yang}},
  \bibinfo{journal}{International Reviews in Physical Chemistry}
  \textbf{\bibinfo{volume}{29}}, \bibinfo{pages}{665} (\bibinfo{year}{2010}).

\bibitem[{\citenamefont{Soler et~al.}(2002)\citenamefont{Soler, Artacho, Gale,
  Garc{\'\i}a, Junquera, Ordej{\'o}n, and S{\'a}nchez-Portal}}]{02JPCM-Josem}
\bibinfo{author}{\bibfnamefont{J.}~\bibnamefont{Soler}},
  \bibinfo{author}{\bibfnamefont{E.}~\bibnamefont{Artacho}},
  \bibinfo{author}{\bibfnamefont{J.~D.} \bibnamefont{Gale}},
  \bibinfo{author}{\bibfnamefont{A.}~\bibnamefont{Garc{\'\i}a}},
  \bibinfo{author}{\bibfnamefont{J.}~\bibnamefont{Junquera}},
  \bibinfo{author}{\bibfnamefont{P.}~\bibnamefont{Ordej{\'o}n}},
  \bibnamefont{and}
  \bibinfo{author}{\bibfnamefont{D.}~\bibnamefont{S{\'a}nchez-Portal}},
  \bibinfo{journal}{Journal of Physics: Condensed Matter}
  \textbf{\bibinfo{volume}{14}}, \bibinfo{pages}{2745} (\bibinfo{year}{2002}).

\bibitem[{\citenamefont{Blum et~al.}(2009)\citenamefont{Blum, Gehrke, Hanke,
  Havu, Havu, Ren, Reuter, and Scheffler}}]{09CPC-Blum}
\bibinfo{author}{\bibfnamefont{V.}~\bibnamefont{Blum}},
  \bibinfo{author}{\bibfnamefont{R.}~\bibnamefont{Gehrke}},
  \bibinfo{author}{\bibfnamefont{F.}~\bibnamefont{Hanke}},
  \bibinfo{author}{\bibfnamefont{P.}~\bibnamefont{Havu}},
  \bibinfo{author}{\bibfnamefont{V.}~\bibnamefont{Havu}},
  \bibinfo{author}{\bibfnamefont{X.}~\bibnamefont{Ren}},
  \bibinfo{author}{\bibfnamefont{K.}~\bibnamefont{Reuter}}, \bibnamefont{and}
  \bibinfo{author}{\bibfnamefont{M.}~\bibnamefont{Scheffler}},
  \bibinfo{journal}{Computer Physics Communications}
  \textbf{\bibinfo{volume}{180}}, \bibinfo{pages}{2175} (\bibinfo{year}{2009}).

\bibitem[{\citenamefont{Ozaki}(2003)}]{03B-Ozaki}
\bibinfo{author}{\bibfnamefont{T.}~\bibnamefont{Ozaki}},
  \bibinfo{journal}{Physical Review B} \textbf{\bibinfo{volume}{67}},
  \bibinfo{pages}{155108} (\bibinfo{year}{2003}).

\bibitem[{\citenamefont{Meng and Kaxiras}(2008)}]{08JCP-Meng}
\bibinfo{author}{\bibfnamefont{S.}~\bibnamefont{Meng}} \bibnamefont{and}
  \bibinfo{author}{\bibfnamefont{E.}~\bibnamefont{Kaxiras}},
  \bibinfo{journal}{The Journal of Chemical Physics}
  \textbf{\bibinfo{volume}{129}}, \bibinfo{pages}{054110}
  (\bibinfo{year}{2008}).

\bibitem[{\citenamefont{Goedecker}(1999)}]{99RMP-Stefan}
\bibinfo{author}{\bibfnamefont{S.}~\bibnamefont{Goedecker}},
  \bibinfo{journal}{Reviews of Modern Physics} \textbf{\bibinfo{volume}{71}},
  \bibinfo{pages}{1085} (\bibinfo{year}{1999}).

\bibitem[{\citenamefont{Bowler and Miyazaki}(2012)}]{12RPP-Bowler}
\bibinfo{author}{\bibfnamefont{D.~R.} \bibnamefont{Bowler}} \bibnamefont{and}
  \bibinfo{author}{\bibfnamefont{T.}~\bibnamefont{Miyazaki}},
  \bibinfo{journal}{Reports on Progress in Physics}
  \textbf{\bibinfo{volume}{75}}, \bibinfo{pages}{036503}
  (\bibinfo{year}{2012}).

\bibitem[{\citenamefont{Pulay}(1969)}]{69MP-Pulay}
\bibinfo{author}{\bibfnamefont{P.}~\bibnamefont{Pulay}},
  \bibinfo{journal}{Molecular Physics} \textbf{\bibinfo{volume}{17}},
  \bibinfo{pages}{197} (\bibinfo{year}{1969}).

\bibitem[{\citenamefont{Li et~al.}(2016)\citenamefont{Li, Liu, Chen, Lin, Ren,
  Lin, Yang, and He}}]{16CMS-Li}
\bibinfo{author}{\bibfnamefont{P.}~\bibnamefont{Li}},
  \bibinfo{author}{\bibfnamefont{X.}~\bibnamefont{Liu}},
  \bibinfo{author}{\bibfnamefont{M.}~\bibnamefont{Chen}},
  \bibinfo{author}{\bibfnamefont{P.}~\bibnamefont{Lin}},
  \bibinfo{author}{\bibfnamefont{X.}~\bibnamefont{Ren}},
  \bibinfo{author}{\bibfnamefont{L.}~\bibnamefont{Lin}},
  \bibinfo{author}{\bibfnamefont{C.}~\bibnamefont{Yang}}, \bibnamefont{and}
  \bibinfo{author}{\bibfnamefont{L.}~\bibnamefont{He}},
  \bibinfo{journal}{Computational Materials Science}
  \textbf{\bibinfo{volume}{112}}, \bibinfo{pages}{503} (\bibinfo{year}{2016}).

\bibitem[{\citenamefont{Arbuznikov and
  Kaupp}(2003{\natexlab{a}})}]{arbuznikov2003self}
\bibinfo{author}{\bibfnamefont{A.~V.} \bibnamefont{Arbuznikov}}
  \bibnamefont{and} \bibinfo{author}{\bibfnamefont{M.}~\bibnamefont{Kaupp}},
  \bibinfo{journal}{Chemical physics letters} \textbf{\bibinfo{volume}{381}},
  \bibinfo{pages}{495} (\bibinfo{year}{2003}{\natexlab{a}}).

\bibitem[{\citenamefont{Skylaris et~al.}(2005)\citenamefont{Skylaris, Haynes,
  Mostofi, and Payne}}]{05JCP-Chris}
\bibinfo{author}{\bibfnamefont{C.-K.} \bibnamefont{Skylaris}},
  \bibinfo{author}{\bibfnamefont{P.~D.} \bibnamefont{Haynes}},
  \bibinfo{author}{\bibfnamefont{A.~A.} \bibnamefont{Mostofi}},
  \bibnamefont{and} \bibinfo{author}{\bibfnamefont{M.~C.} \bibnamefont{Payne}},
  \bibinfo{journal}{The Journal of Chemical Physics}
  \textbf{\bibinfo{volume}{122}}, \bibinfo{pages}{084119}
  (\bibinfo{year}{2005}).

\bibitem[{\citenamefont{Perdew et~al.}(2009)\citenamefont{Perdew, Ruzsinszky,
  Csonka, Constantin, and Sun}}]{09L-Perdew}
\bibinfo{author}{\bibfnamefont{J.~P.} \bibnamefont{Perdew}},
  \bibinfo{author}{\bibfnamefont{A.}~\bibnamefont{Ruzsinszky}},
  \bibinfo{author}{\bibfnamefont{G.~I.} \bibnamefont{Csonka}},
  \bibinfo{author}{\bibfnamefont{L.~A.} \bibnamefont{Constantin}},
  \bibnamefont{and} \bibinfo{author}{\bibfnamefont{J.}~\bibnamefont{Sun}},
  \bibinfo{journal}{Physical Review Letters} \textbf{\bibinfo{volume}{103}},
  \bibinfo{pages}{026403} (\bibinfo{year}{2009}).

\bibitem[{\citenamefont{Bart{\'o}k and
  Yates}(2019{\natexlab{a}})}]{19JCP-Bartok}
\bibinfo{author}{\bibfnamefont{A.~P.} \bibnamefont{Bart{\'o}k}}
  \bibnamefont{and} \bibinfo{author}{\bibfnamefont{J.~R.} \bibnamefont{Yates}},
  \bibinfo{journal}{The Journal of Chemical Physics}
  \textbf{\bibinfo{volume}{150}}, \bibinfo{pages}{161101}
  (\bibinfo{year}{2019}{\natexlab{a}}).

\bibitem[{\citenamefont{Lehtola et~al.}(2018)\citenamefont{Lehtola, Steigemann,
  Oliveira, and Marques}}]{18SX-Lehtola}
\bibinfo{author}{\bibfnamefont{S.}~\bibnamefont{Lehtola}},
  \bibinfo{author}{\bibfnamefont{C.}~\bibnamefont{Steigemann}},
  \bibinfo{author}{\bibfnamefont{M.~J.~T.} \bibnamefont{Oliveira}},
  \bibnamefont{and} \bibinfo{author}{\bibfnamefont{M.~A.~L.}
  \bibnamefont{Marques}}, \bibinfo{journal}{SoftwareX}
  \textbf{\bibinfo{volume}{7}}, \bibinfo{pages}{1} (\bibinfo{year}{2018}).

\bibitem[{\citenamefont{Zahariev et~al.}(2013)\citenamefont{Zahariev, Leang,
  and Gordon}}]{13JCP-Zahariev}
\bibinfo{author}{\bibfnamefont{F.}~\bibnamefont{Zahariev}},
  \bibinfo{author}{\bibfnamefont{S.~S.} \bibnamefont{Leang}}, \bibnamefont{and}
  \bibinfo{author}{\bibfnamefont{M.~S.} \bibnamefont{Gordon}},
  \bibinfo{journal}{The Journal of Chemical Physics}
  \textbf{\bibinfo{volume}{138}}, \bibinfo{pages}{244108}
  (\bibinfo{year}{2013}).

\bibitem[{\citenamefont{Yang et~al.}(2016)\citenamefont{Yang, Peng, Sun, and
  Perdew}}]{yang2016more}
\bibinfo{author}{\bibfnamefont{Z.-h.} \bibnamefont{Yang}},
  \bibinfo{author}{\bibfnamefont{H.}~\bibnamefont{Peng}},
  \bibinfo{author}{\bibfnamefont{J.}~\bibnamefont{Sun}}, \bibnamefont{and}
  \bibinfo{author}{\bibfnamefont{J.~P.} \bibnamefont{Perdew}},
  \bibinfo{journal}{Physical review B} \textbf{\bibinfo{volume}{93}},
  \bibinfo{pages}{205205} (\bibinfo{year}{2016}).

\bibitem[{\citenamefont{Arbuznikov and
  Kaupp}(2003{\natexlab{b}})}]{03CPL-Alexei}
\bibinfo{author}{\bibfnamefont{A.~V.} \bibnamefont{Arbuznikov}}
  \bibnamefont{and} \bibinfo{author}{\bibfnamefont{M.}~\bibnamefont{Kaupp}},
  \bibinfo{journal}{Chemical Physics Letters} \textbf{\bibinfo{volume}{381}},
  \bibinfo{pages}{495} (\bibinfo{year}{2003}{\natexlab{b}}).

\bibitem[{\citenamefont{Becke and Roussel}(1989)}]{89A-Becke}
\bibinfo{author}{\bibfnamefont{A.~D.} \bibnamefont{Becke}} \bibnamefont{and}
  \bibinfo{author}{\bibfnamefont{M.~R.} \bibnamefont{Roussel}},
  \bibinfo{journal}{Physical Review A} \textbf{\bibinfo{volume}{39}},
  \bibinfo{pages}{3761} (\bibinfo{year}{1989}).

\bibitem[{\citenamefont{Van~Voorhis and Scuseria}(1998)}]{98JCP-Troy}
\bibinfo{author}{\bibfnamefont{T.}~\bibnamefont{Van~Voorhis}} \bibnamefont{and}
  \bibinfo{author}{\bibfnamefont{G.~E.} \bibnamefont{Scuseria}},
  \bibinfo{journal}{The Journal of Chemical Physics}
  \textbf{\bibinfo{volume}{109}}, \bibinfo{pages}{400} (\bibinfo{year}{1998}).

\bibitem[{\citenamefont{Arbuznikov et~al.}(2002)\citenamefont{Arbuznikov,
  Kaupp, Malkin, Reviakine, and Malkina}}]{02PCCP-Alexei}
\bibinfo{author}{\bibfnamefont{A.~V.} \bibnamefont{Arbuznikov}},
  \bibinfo{author}{\bibfnamefont{M.}~\bibnamefont{Kaupp}},
  \bibinfo{author}{\bibfnamefont{V.~G.} \bibnamefont{Malkin}},
  \bibinfo{author}{\bibfnamefont{R.}~\bibnamefont{Reviakine}},
  \bibnamefont{and} \bibinfo{author}{\bibfnamefont{O.~L.}
  \bibnamefont{Malkina}}, \bibinfo{journal}{Physical Chemistry Chemical
  Physics} \textbf{\bibinfo{volume}{4}}, \bibinfo{pages}{5467}
  (\bibinfo{year}{2002}).

\bibitem[{\citenamefont{Zheng et~al.}(2021)\citenamefont{Zheng, Ren, and
  He}}]{21CPC-Zheng}
\bibinfo{author}{\bibfnamefont{D.}~\bibnamefont{Zheng}},
  \bibinfo{author}{\bibfnamefont{X.}~\bibnamefont{Ren}}, \bibnamefont{and}
  \bibinfo{author}{\bibfnamefont{L.}~\bibnamefont{He}},
  \bibinfo{journal}{Computer Physics Communications}
  \textbf{\bibinfo{volume}{267}}, \bibinfo{pages}{108043}
  (\bibinfo{year}{2021}).

\bibitem[{\citenamefont{Anderson et~al.}(1999)\citenamefont{Anderson, Bai,
  Bischof, Blackford, Demmel, Dongarra, Croz, Greenbaum, Hammarling, McKenney
  et~al.}}]{99-Anderson}
\bibinfo{author}{\bibfnamefont{E.}~\bibnamefont{Anderson}},
  \bibinfo{author}{\bibfnamefont{Z.}~\bibnamefont{Bai}},
  \bibinfo{author}{\bibfnamefont{C.}~\bibnamefont{Bischof}},
  \bibinfo{author}{\bibfnamefont{S.}~\bibnamefont{Blackford}},
  \bibinfo{author}{\bibfnamefont{J.}~\bibnamefont{Demmel}},
  \bibinfo{author}{\bibfnamefont{J.}~\bibnamefont{Dongarra}},
  \bibinfo{author}{\bibfnamefont{J.~D.} \bibnamefont{Croz}},
  \bibinfo{author}{\bibfnamefont{A.}~\bibnamefont{Greenbaum}},
  \bibinfo{author}{\bibfnamefont{S.}~\bibnamefont{Hammarling}},
  \bibinfo{author}{\bibfnamefont{A.}~\bibnamefont{McKenney}},
  \bibnamefont{et~al.}, \emph{\bibinfo{title}{LAPACK Users' Guide}}
  (\bibinfo{publisher}{Society of Industrial and Applied Mathematics},
  \bibinfo{year}{1999}).

\bibitem[{\citenamefont{Peng et~al.}(2016)\citenamefont{Peng, Yang, Perdew, and
  Sun}}]{16X-Peng}
\bibinfo{author}{\bibfnamefont{H.}~\bibnamefont{Peng}},
  \bibinfo{author}{\bibfnamefont{Z.-H.} \bibnamefont{Yang}},
  \bibinfo{author}{\bibfnamefont{J.~P.} \bibnamefont{Perdew}},
  \bibnamefont{and} \bibinfo{author}{\bibfnamefont{J.}~\bibnamefont{Sun}},
  \bibinfo{journal}{Physical Review X} \textbf{\bibinfo{volume}{6}},
  \bibinfo{pages}{041005} (\bibinfo{year}{2016}).

\bibitem[{\citenamefont{Jure{\v c}ka et~al.}(2006)\citenamefont{Jure{\v c}ka,
  {\v S}poner, {\v C}ern{\'y}, and Hobza}}]{06PCCP-Petr}
\bibinfo{author}{\bibfnamefont{P.}~\bibnamefont{Jure{\v c}ka}},
  \bibinfo{author}{\bibfnamefont{J.}~\bibnamefont{{\v S}poner}},
  \bibinfo{author}{\bibfnamefont{J.}~\bibnamefont{{\v C}ern{\'y}}},
  \bibnamefont{and} \bibinfo{author}{\bibfnamefont{P.}~\bibnamefont{Hobza}},
  \bibinfo{journal}{Physical Chemistry Chemical Physics}
  \textbf{\bibinfo{volume}{8}}, \bibinfo{pages}{1985} (\bibinfo{year}{2006}).

\bibitem[{\citenamefont{Schlipf and Gygi}(2015)}]{15CPC-Schlipf}
\bibinfo{author}{\bibfnamefont{M.}~\bibnamefont{Schlipf}} \bibnamefont{and}
  \bibinfo{author}{\bibfnamefont{F.}~\bibnamefont{Gygi}},
  \bibinfo{journal}{Computer Physics Communications}
  \textbf{\bibinfo{volume}{196}}, \bibinfo{pages}{36} (\bibinfo{year}{2015}).

\bibitem[{\citenamefont{Perdew et~al.}(1996)\citenamefont{Perdew, Burke, and
  Ernzerhof}}]{96L-Perdew}
\bibinfo{author}{\bibfnamefont{J.~P.} \bibnamefont{Perdew}},
  \bibinfo{author}{\bibfnamefont{K.}~\bibnamefont{Burke}}, \bibnamefont{and}
  \bibinfo{author}{\bibfnamefont{M.}~\bibnamefont{Ernzerhof}},
  \bibinfo{journal}{Phys. Rev. Lett.} \textbf{\bibinfo{volume}{77}},
  \bibinfo{pages}{3865} (\bibinfo{year}{1996}).

\bibitem[{\citenamefont{Caro et~al.}(2012)\citenamefont{Caro, Schulz, and
  O~Reilly}}]{caro2012comparison}
\bibinfo{author}{\bibfnamefont{M.~A.} \bibnamefont{Caro}},
  \bibinfo{author}{\bibfnamefont{S.}~\bibnamefont{Schulz}}, \bibnamefont{and}
  \bibinfo{author}{\bibfnamefont{E.~P.} \bibnamefont{O~Reilly}},
  \bibinfo{journal}{Journal of Physics: Condensed Matter}
  \textbf{\bibinfo{volume}{25}}, \bibinfo{pages}{025803}
  (\bibinfo{year}{2012}).

\bibitem[{\citenamefont{Furness et~al.}(2020)\citenamefont{Furness, Kaplan,
  Ning, Perdew, and Sun}}]{20JPCL-Furness}
\bibinfo{author}{\bibfnamefont{J.~W.} \bibnamefont{Furness}},
  \bibinfo{author}{\bibfnamefont{A.~D.} \bibnamefont{Kaplan}},
  \bibinfo{author}{\bibfnamefont{J.}~\bibnamefont{Ning}},
  \bibinfo{author}{\bibfnamefont{J.~P.} \bibnamefont{Perdew}},
  \bibnamefont{and} \bibinfo{author}{\bibfnamefont{J.}~\bibnamefont{Sun}},
  \bibinfo{journal}{The Journal of Physical Chemistry Letters}
  \textbf{\bibinfo{volume}{11}}, \bibinfo{pages}{8208} (\bibinfo{year}{2020}).

\bibitem[{\citenamefont{Bart{\'o}k and
  Yates}(2019{\natexlab{b}})}]{bartok2019regularized}
\bibinfo{author}{\bibfnamefont{A.~P.} \bibnamefont{Bart{\'o}k}}
  \bibnamefont{and} \bibinfo{author}{\bibfnamefont{J.~R.} \bibnamefont{Yates}},
  \bibinfo{journal}{The Journal of chemical physics}
  \textbf{\bibinfo{volume}{150}}, \bibinfo{pages}{161101}
  (\bibinfo{year}{2019}{\natexlab{b}}).

\bibitem[{\citenamefont{DiStasio et~al.}(2014)\citenamefont{DiStasio, Santra,
  Li, Wu, and Car}}]{14JCP-Distasio}
\bibinfo{author}{\bibfnamefont{R.~A.} \bibnamefont{DiStasio}},
  \bibinfo{author}{\bibfnamefont{B.}~\bibnamefont{Santra}},
  \bibinfo{author}{\bibfnamefont{Z.}~\bibnamefont{Li}},
  \bibinfo{author}{\bibfnamefont{X.}~\bibnamefont{Wu}}, \bibnamefont{and}
  \bibinfo{author}{\bibfnamefont{R.}~\bibnamefont{Car}}, \bibinfo{journal}{The
  Journal of Chemical Physics} \textbf{\bibinfo{volume}{141}},
  \bibinfo{pages}{084502} (\bibinfo{year}{2014}).

\bibitem[{\citenamefont{Santra et~al.}(2008)\citenamefont{Santra, Michaelides,
  Fuchs, Tkatchenko, Filippi, and Scheffler}}]{08JCP-Santra}
\bibinfo{author}{\bibfnamefont{B.}~\bibnamefont{Santra}},
  \bibinfo{author}{\bibfnamefont{A.}~\bibnamefont{Michaelides}},
  \bibinfo{author}{\bibfnamefont{M.}~\bibnamefont{Fuchs}},
  \bibinfo{author}{\bibfnamefont{A.}~\bibnamefont{Tkatchenko}},
  \bibinfo{author}{\bibfnamefont{C.}~\bibnamefont{Filippi}}, \bibnamefont{and}
  \bibinfo{author}{\bibfnamefont{M.}~\bibnamefont{Scheffler}},
  \bibinfo{journal}{The Journal of Chemical Physics}
  \textbf{\bibinfo{volume}{129}}, \bibinfo{pages}{194111}
  (\bibinfo{year}{2008}).

\bibitem[{\citenamefont{Kim and Kim}(1998)}]{98JCP-Kim}
\bibinfo{author}{\bibfnamefont{J.}~\bibnamefont{Kim}} \bibnamefont{and}
  \bibinfo{author}{\bibfnamefont{K.~S.} \bibnamefont{Kim}},
  \bibinfo{journal}{The Journal of Chemical Physics}
  \textbf{\bibinfo{volume}{109}}, \bibinfo{pages}{5886} (\bibinfo{year}{1998}).

\bibitem[{\citenamefont{Losada and Leutwyler}(2002)}]{02JCP-Losada}
\bibinfo{author}{\bibfnamefont{M.}~\bibnamefont{Losada}} \bibnamefont{and}
  \bibinfo{author}{\bibfnamefont{S.}~\bibnamefont{Leutwyler}},
  \bibinfo{journal}{The Journal of Chemical Physics}
  \textbf{\bibinfo{volume}{117}}, \bibinfo{pages}{2003} (\bibinfo{year}{2002}).

\bibitem[{\citenamefont{Podeszwa et~al.}(2010)\citenamefont{Podeszwa,
  Patkowski, and Szalewicz}}]{10PCCP-Podeszwa}
\bibinfo{author}{\bibfnamefont{R.}~\bibnamefont{Podeszwa}},
  \bibinfo{author}{\bibfnamefont{K.}~\bibnamefont{Patkowski}},
  \bibnamefont{and}
  \bibinfo{author}{\bibfnamefont{K.}~\bibnamefont{Szalewicz}},
  \bibinfo{journal}{Physical Chemistry Chemical Physics}
  \textbf{\bibinfo{volume}{12}}, \bibinfo{pages}{5974} (\bibinfo{year}{2010}).

\bibitem[{\citenamefont{Harl et~al.}(2010)\citenamefont{Harl, Schimka, and
  Kresse}}]{10B-Harl}
\bibinfo{author}{\bibfnamefont{J.}~\bibnamefont{Harl}},
  \bibinfo{author}{\bibfnamefont{L.}~\bibnamefont{Schimka}}, \bibnamefont{and}
  \bibinfo{author}{\bibfnamefont{G.}~\bibnamefont{Kresse}},
  \bibinfo{journal}{Physical Review B} \textbf{\bibinfo{volume}{81}},
  \bibinfo{pages}{115126} (\bibinfo{year}{2010}).

\bibitem[{\citenamefont{M\o{}ller and Plesset}(1934)}]{34PR-Moller}
\bibinfo{author}{\bibfnamefont{C.}~\bibnamefont{M\o{}ller}} \bibnamefont{and}
  \bibinfo{author}{\bibfnamefont{M.~S.} \bibnamefont{Plesset}},
  \bibinfo{journal}{Physical Review} \textbf{\bibinfo{volume}{46}},
  \bibinfo{pages}{618} (\bibinfo{year}{1934}).

\bibitem[{\citenamefont{Boys and Bernardi}(1970)}]{70MP-Boys}
\bibinfo{author}{\bibfnamefont{S.~F.} \bibnamefont{Boys}} \bibnamefont{and}
  \bibinfo{author}{\bibfnamefont{F.}~\bibnamefont{Bernardi}},
  \bibinfo{journal}{Molecular Physics} \textbf{\bibinfo{volume}{19}},
  \bibinfo{pages}{553} (\bibinfo{year}{1970}).

\bibitem[{\citenamefont{Bl\"ochl}(1994)}]{94B-Blochl}
\bibinfo{author}{\bibfnamefont{P.~E.} \bibnamefont{Bl\"ochl}},
  \bibinfo{journal}{Physical Review B} \textbf{\bibinfo{volume}{50}},
  \bibinfo{pages}{17953} (\bibinfo{year}{1994}).

\end{thebibliography}
\end{document}